\documentclass[12pt]{article}
\usepackage[a4paper,margin=2.5cm]{geometry} 
\usepackage{graphicx}  
\usepackage{hyperref}  
\usepackage{natbib}    
\usepackage{url}       
\usepackage{setspace, lineno}
\usepackage{amsmath, amssymb, amsfonts}
\usepackage{bbm}
\usepackage{comment}
\usepackage{algorithm}
\usepackage{algorithmic}
\usepackage[table]{xcolor}
\usepackage{hyperref}

\usepackage{booktabs}
\usepackage{siunitx}
\title{\textbf{The Last Expectation: Sampling Measurement Outcomes in Value of Information Analysis}}

\author{Lea Friedli\footnote{Engineering Risk Analysis Group, Technical University of Munich, Germany} \footnote{Corresponding author: Theresienstr. 90, 80333 Munich, Germany, lea.friedli@tum.de}, Iason Papaioannou\footnote{Laboratory of Structural Analysis and Antiseismic Research, School of Civil Engineering, National Technical University of Athens, Greece} and Daniel Straub$^*$ \\
}
\date{ }

\newcommand \X {\boldsymbol{X}}%
\newcommand \x {\boldsymbol{x}}%
\newcommand \Y {\boldsymbol{Y}}%
\newcommand \y {\boldsymbol{y}}%
\newcommand \RXa {\mathcal{R}\left(\boldsymbol{X},a\right)}%
\newcommand \Rxa {\mathcal{R}\left(\boldsymbol{x},a\right)}%
\newcommand \GX {\mathcal{G}\left(\boldsymbol{X}\right)}%
\newcommand \Gx {\mathcal{G}\left(\boldsymbol{x}\right)}%
\newcommand \thres {T}

\newcommand \noise {\boldsymbol{r}}
\newcommand \Noise {\boldsymbol{R}}
\newcommand \aprior {a_{\mathrm{prior}}}

\begin{document}

\maketitle

\begin{abstract}
    In decision-making under uncertainty, additional information can improve decisions but is often costly to acquire. The Value of Information (VoI) quantifies the expected benefit of different information-gathering strategies. The main computational burden in VoI analysis arises from evaluating the expected utility of a strategy over possible outcomes. As each outcome entails solving a computationally expensive optimization problem under uncertainty to determine the optimal decision, efficient estimation of this expectation is crucial. This work focuses on sampling-based estimation and investigates three classical variance-reduction techniques, namely control variates, stratified sampling, and importance sampling, as well as combinations thereof. In particular, we establish a connection between two common VoI formulations through a control variate framework and show how this relationship can be exploited to obtain more efficient VoI estimators.
    The methodology is assessed using three test cases. The first is an analytical example with a known reference solution, the second involves a fatigue reliability and the third a slope stability application. Our results demonstrate that the relative performance of the sampling methods depends strongly on the problem characteristics. However, suitable combinations of variance-reduction techniques consistently improve efficiency and provide stable VoI estimates across a range of scenarios.
\end{abstract}

\section{Introduction}

Decision analysis entails applying principles from decision theory to make well-informed decisions \citep{raiffa1961applied, howard1966information}. Decision-makers navigate uncertainty, balancing the costs of actions against their associated risks and consequences. In practice, many decisions rely on uncertain parameters, such as soil properties in geotechnical engineering or the condition of machine components in mechanical engineering. While obtaining additional information on these uncertain parameters has the potential to decrease uncertainty and enhance decision outcomes, it often entails high costs. Hence, information acquisition strategies should be chosen carefully, a problem commonly addressed within the framework of pre-posterior decision analysis, where the expected benefit of acquiring information is evaluated before outcomes become available.  \\

The Value of Information (VoI), rooted in Bayesian statistical decision theory \citep{raiffa1961applied, howard1966information, degroot2005optimal}, serves to evaluate the advantages of acquiring additional observations for better informed decisions. 
It is also applied in sensitivity analysis by assigning decision-relevant metrics to reductions in uncertainty \citep[e.g.,][]{porn1997decision, oakley2009decision, borgonovo2017deciding, straub2022decision}.
In the realm of engineering applications, the VoI finds widespread use in quantifying the influence of monitoring and inspections on structural systems \citep[e.g.,][]{straub2005risk,pozzi2011, malings2016value, thons2018value, luque2019risk, vereecken2020bayesian, kamariotis2022value, giordano2023value}. Also in the field of geosciences, the applications of the VoI are diverse, extending to various areas such as geological exploration \citep[e.g.,][]{eidsvik2008value, athens2019monte, dutta2019value} and groundwater management \citep[e.g.,][]{liu2012value, trainor2014geophysical}. For a recent review of VoI in civil and infrastructure engineering, see \citet{zhang2021value}. \\

We adopt the standard expected utility framework, in which the optimal decision under uncertainty is the one that maximizes expected utility \citep{von2007theory}. For simplicity, we assume that all consequences can be expressed as monetary costs and that utility is linear in cost. However, the findings of this work remain valid for more general utility functions. Here, the optimal decision is the one that minimizes the expected cost. Furthermore, we focus on measurements in a continuous space as the source of information.
The resulting VoI of a measurement strategy is the expected reduction in cost that can be achieved by making decisions accounting for the additional information provided by the measurements. The optimal measurement system maximizes the net value of information, i.e., the VoI minus the measurement cost \citep{straub2014value}.\\

In practice, measurement systems provide only imperfect information about the underlying uncertain quantities. Bayesian updating is used to incorporate such information and revise the probability distributions of the underlying variables \citep{beck1998updating, straub2015bayesian}.
For a given measurement outcome, evaluating the minimum expected cost requires identifying the optimal action, which in turn involves assessing the system performance under the candidate actions, often through expensive reliability analyses \citep[e.g., ][]{dutta2019value,ehre2021conditional,jiang2020optimization}.
Depending on whether the action space is continuous or discrete, the optimization problem may be non-convex or combinatorial, leading to additional challenges \citep{malings2016value,dutta2019value}.
The central challenge in VoI analysis, however, is that the outcome remains unknown when a measurement strategy is selected. Consequently, the benefit of the strategy must be averaged over all possible measurement outcomes. This outer expectation, which accounts for the uncertainty in future measurements, is referred to in the following as the ``last expectation''. \\

In practical problems, this last expectation does not have a closed-form expression and must be approximated numerically. Although  numerical integration may be feasible in some settings \citep[e.g.,][]{heath2016estimating, konakli2016numerical}, sampling-based methods are generally more flexible, particularly in high-dimensional parameter settings. Estimation of the last expectation requires a computationally expensive cost-minimization analysis for each sampled measurement outcome.
Hence, the overall efficiency of VoI evaluation depends critically on the sampling strategy employed.
The most widely used approach is Monte Carlo (MC) sampling  \citep{eidsvik2008value, dutta2019value, vereecken2020bayesian, kamariotis2022value}. However, MC sampling is inefficient when measurement outcomes that lead to changes in the optimal decisions occur only rarely \citep{straub2014value}. This issue becomes more pronounced when multiple measurements with several possible outcomes are considered. \\

This work investigates and combines classical variance-reduction techniques \citep[e.g.,][]{rubinstein2017simulation} for efficient VoI estimation, namely control variates, stratified sampling, and importance sampling. To the best of our knowledge, the numerical estimation of the last expectation in VoI analysis has not previously been the subject of a systematic investigation.
In particular, we show how two alternative VoI formulations are related through a control variate framework, and demonstrate that the resulting control variate estimator can provide improved robustness and efficiency. Furthermore, particular attention is given to importance sampling, for which a suitable proposal density must be constructed. To this end, we use Gaussian process (GP) models to approximate the optimal IS density. This approach is inspired by the work of \citet{dubourg2013metamodel} and \citet{dalbey2014gaussian}, who similarly combine GP surrogates and importance sampling, albeit in the context of rare-event probability estimation rather than VoI analysis. We investigate the proposed methods on three test cases, all based on a binary decision between taking no action and a perfect action, where the latter completely eliminates the failure risk. 
The first test case is an analytical example that enables benchmarking against a known reference solution. The second and third test cases are adapted from \citet{kanjilal2023bayesian} and involve fatigue crack growth and slope stability, respectively. \\

The remainder of this manuscript is structured as follows: Section~2 presents the theoretical background, including simple decision-making, Bayesian updating, and the concept of the value of information. In Section~3, the estimation of the last expectation is discussed, covering Monte Carlo sampling, control variates, stratified sampling, and importance sampling. Section~4 provides numerical illustrations and Section~5 discusses the results and outlines future research directions. Finally, Section~6 concludes the manuscript.

\section{Background}
\label{back}

\subsection{Simple decision}
\label{sec:simple}

Our focus is on simple decision-making scenarios, where a single choice must be made in an uncertain environment \citep[see][]{kochenderfer2022algorithms}. 
In the considered decision scenario, we represent the uncertain state of the world by the random vector~$\X$ with outcomes in~$\mathbb{R}^d$ . We denote the prior probability density function (PDF) of $\X$ by $f_{\X}$. 
The decision evolves around a potential failure of the corresponding system and we are considering actions $a \in \mathcal{A}$ (continuous or discrete) which can be undertaken to prevent system failure. 
The system performance is modeled by a non-linear function $\mathcal{R}: \mathbb{R}^d \times \mathcal{A} \rightarrow \mathbb{R}$, which maps a given action $a$ and state of the world $\x$ to a quantity of interest, e.g., related to the system stress state or deformation. System failure (binary system state) occurs if $\RXa \geq \thres$, with $\thres \in \mathbb{R}$ denoting a critical threshold. If the cost of action $a$ is $c_A(a)$ and the cost of failure is $c_F$, we can express the expected cost for action~$a$ as:
\begin{align*}
    \mathbb{E}[C | a] = \int \left( c_A(a) + c_F \; \mathbbm{I} \{ \Rxa \geq \thres \} \right) f_{\X}(\x) \; \mathrm{d} \x
    = c_A(a) + c_F \; p_F(a),
\end{align*}
where $\mathbbm{I}(\cdot)$ is the indicator function and, 
$$p_F(a) = \mathbb{P} \left( \RXa \geq \thres \right) = \int \mathbbm{I} \{ \Rxa \geq \thres \} f_{\X}(\x) \; \mathrm{d} \x,$$ 
is the probability of failure given action $a$. Although we focus on this setting, many findings of this work remain valid for more general settings.

\subsection{Bayesian updating}
\label{EBM_ebm_setting}

We are considering additional information about $\X$ given by the measurements~$\y \in \mathbb{R}^m$. In practice, measurement systems provide only imperfect information, and measurements are typically affected by uncertainty and random errors. Bayesian updating provides a framework for incorporating such imperfect information to update the probability distribution of $\X$ \citep{beck1998updating, straub2015bayesian}.
In most applications, the random measurement vector is given by $\Y = \GX + \noise$, where $\mathcal{G}: \mathbb{R}^d \rightarrow \mathbb{R}^m$ is a forward operator and $\noise$ is a random  measurement error with outcomes in~$\mathbb{R}^m$. We rely on Bayes' theorem and obtain the posterior PDF of $\X$ given the measurements $\y$ as: 
\begin{align*}
f_{\X|\Y}(\x | \y ) = \frac{f_{\X}(\x) f_{\Y|\X}(\y | \x) }{f_{\Y}(\y)},
\end{align*}
with the prior PDF~$f_{\X}(\x)$, the likelihood function~$f_{\Y|\X}(\y|\x)$ and the evidence~$f_{\Y}(\y)$ (assumed non-zero). 
Assuming Gaussian measurement errors, the likelihood is expressed as $f_{\Y|\X}(\y|\x)=\varphi_{m}(\y; \Gx, \boldsymbol{\Sigma_{\Y}})$, with $\varphi_{m}(\cdot; \boldsymbol{\mu}, \boldsymbol{\Sigma})$ denoting the PDF of a $m$-variate normal distribution with mean $\boldsymbol{\mu}$ and covariance matrix~$\boldsymbol{\Sigma}$. If the errors are independent, $\boldsymbol{\Sigma_{\Y}}$ is given by a diagonal matrix with the variance of the measurement error on its diagonal. \\

Using this posterior PDF, we can update the probability of failure for a given action $a$:
\begin{equation}
    \label{prob_fail_up}
    p_F(a | \y) = \mathbb{P} \left( \RXa \geq \thres | \y \right) = \int \mathbbm{I} \{ \Rxa \geq \thres \} f_{\X|\Y}(\x|\y) \mathrm{d} \x.
\end{equation}
This leads to the updated expected cost 
$\mathbb{E}[C | a, \y] = c_A(a) + c_F \; p_F(a | \y )$.

\subsection{Value of Information}
Having defined the expected cost given a specific action and measurement combination, we can now move on to define the~Value of Information~(VoI).
The VoI is the expected reduction in decision cost that can be achieved by acquiring additional information before making a decision.
Let $\aprior = \underset{a}{\mathrm{argmin}} \; \mathbb{E}[C | a]$ denote the a-priori optimal decision, minimizing the expected cost before any additional information is acquired. We consider the following mathematically equivalent expressions of the VoI, which are both used in the literature \citep[e.g.,][]{straub2014value,jiang2020optimization}:
\begin{align}
    \label{VoIa}
    \text{VoI}
    &= \int f_{\Y}(\y) \; \left( \mathbb{E}[C | \aprior, \y ] 
    - \underset{a}{\min} \; \mathbb{E}[C| a, \y] \right) \;  d \y, \tag{A} \\
    \label{VoIb}
    &= \mathbb{E}[C | \aprior] - \int f_{\Y}(\y) \; \underset{a}{\min} \; \mathbb{E}[C| a, \y] \;  d \y. \tag{B}
\end{align}
Expression~(\ref{VoIa}) defines the VoI as the expected reduction in optimal cost due to the additional information, obtained by averaging the cost reduction over all possible measurement outcomes. Equivalently, expression~(\ref{VoIb}) expresses the VoI as the difference between the expected cost of the a-priori optimal decision and the expected a-posteriori optimal cost, where the latter is averaged over all possible measurement outcomes.\\

There are three main computational components involved in evaluating the VoI \citep{malings2016value}, which can be observed in Equations \ref{VoIa} and \ref{VoIb}: (1) the computation of the expected cost for an action - measurement combination, $\mathbb{E}[C| a, \y]$, (2) the minimization of the expected cost over possible actions and (3) the outer integration over all possible measurement outcomes $\Y$. This manuscript focuses on the third component and investigates different sampling methods for its efficient evaluation.

\section{The last expectation}
\label{main}

For the approximation of the expectation over $\Y$ in Equations (\ref{VoIa}) and (\ref{VoIb}), we consider MC estimation (Sct.~\ref{sec_mc}), and variance-reduction strategies relying on control variates (Sct.~\ref{sec_cv}), stratified sampling (Sct.~\ref{sec_lhs}) and importance sampling (Sct.~\ref{sec_is_gen}).

\subsection{Monte Carlo sampling}
\label{sec_mc}
Let us define the following functions of $\y$:
\begin{align*}
    h_1(\y) = \mathbb{E}[C \mid \aprior, \y], \quad
    h_B(\y) = \min_{a} \; \mathbb{E}[C \mid a, \y], \quad
    h_A(\y) = h_1(\y) - h_B(\y).
\end{align*}

For Monte Carlo (MC) estimation of the VoI expressions in
Equations~(\ref{VoIa}) and~(\ref{VoIb}), we use independent and identically distributed (i.i.d.) samples
$\{ \y_1, \dots \y_n \} \sim f_{\Y}(\y)$ and:
\begin{align}
    \label{eq:mc-estimators}
    \widehat{\theta}_A^{\mathrm{MC}}(n)
    = \frac{1}{n} \sum_{j=1}^n h_A(\y_j), \quad
    \widehat{\theta}_B^{\mathrm{MC}}(n)
    = \mathbb{E}[C \mid \aprior]
    - \frac{1}{n} \sum_{j=1}^n h_B(\y_j).
\end{align}

Within this manuscript, we assume that $\aprior$ and the corresponding expected cost
$\mathbb{E}[C~\mid~\aprior]$, such as the posterior expected cost $\mathbb{E}[C \mid a, \y]$ given $\y$ can be evaluated exactly or to a sufficiently high degree of accuracy. 
This means that within the setup considered in this study (Sect.~\ref{sec:simple}), we assume that all failure probabilities can be evaluated with negligible uncertainty.
This assumption enables us to isolate the sampling uncertainty due to estimation of the last expectation.
Under this assumption, the variances of the two MC
estimators are
\begin{align*}
    \mathrm{Var}\!\left( \widehat{\theta}_A^{\mathrm{MC}}(n) \right)
    = \frac{1}{n} \, \mathrm{Var}\!\left(h_A(\Y)\right), \quad
    \mathrm{Var}\!\left( \widehat{\theta}_B^{\mathrm{MC}}(n) \right)
    = \frac{1}{n} \, \mathrm{Var}\!\left(h_B(\Y)\right).
\end{align*}

From $\mathrm{Var}\!\left(h_A(\Y)\right)
    = \mathrm{Var}\!\left(h_1(\Y)\right)
    + \mathrm{Var}\!\left(h_B(\Y)\right)
    - 2\,\mathrm{Cov}\!\left(h_1(\Y), h_B(\Y)\right)$ it follows that the MC estimator $\widehat{\theta}_A^{\mathrm{MC}}$ 
has lower variance than $\widehat{\theta}_B^{\mathrm{MC}}$ if $\mathrm{Var}\!\left(h_1(\Y)\right) < 2\,\mathrm{Cov}\!\left(h_1(\Y), h_B(\Y)\right)$, and vice versa. 
This criterion could, in principle, be used to select the more efficient estimator based on estimates of the variance and covariance terms.
However, as shown in the following section, a more effective approach is to combine the two estimators, yielding a lower variance than either estimator individually.

\subsection{Control variate}
\label{sec_cv}
We can form a convex combination of the two MC estimators
(\ref{VoIa}) and (\ref{VoIb}) as
\begin{align}
\label{eq:mc-estimators-comb}
    \widehat{\theta}_{\alpha}^{\mathrm{MC}}(n)
    &= \alpha \, \widehat{\theta}_A^{\mathrm{MC}}(n)
    + (1-\alpha)\, \widehat{\theta}_B^{\mathrm{MC}}(n) \\
    &= (1-\alpha)\,\mathbb{E}[C \mid \aprior]
    + \frac{1}{n} \sum_{j=1}^n
    \left[
        \alpha\, h_1(\y_j) - h_B(\y_j)
    \right]. \nonumber
\end{align}
An appropriate choice of the coefficient $\alpha$ is crucial \citep{robert2004monte}.
Therefore, $\alpha$ is chosen to minimize the variance of the
combined estimator,
\begin{align*}
    \mathrm{Var}\!\left(\widehat{\theta}_{\alpha}^{\mathrm{MC}}(n) \right)
    &=
    \frac{1}{n}
    \Big[
        \alpha^2\, \mathrm{Var}\!\left(h_1(\Y)\right)
        + \mathrm{Var}\!\left(h_B(\Y)\right)
        - 2\alpha \,
        \mathrm{Cov}\!\left(h_1(\Y), h_B(\Y)\right)
    \Big],
\end{align*}
which is a quadratic function of $\alpha$. The minimizing choice is
\begin{equation}
    \label{alpha}
    \alpha^*
    =
    \frac{
        \mathrm{Cov}\!\left(h_1(\Y), h_B(\Y)\right)
    }{
        \mathrm{Var}\!\left(h_1(\Y)\right)
    }.
\end{equation}
This result aligns with the either--or selection criterion discussed in Section \ref{sec_mc}:  \linebreak if
$\mathrm{Var}(h_1(\Y))~\leq~2\,\mathrm{Cov}(h_1(\Y),~h_B(\Y))$,
then $\alpha^* \geq 0.5$, placing more weight on
$\widehat{\theta}_A^{\mathrm{MC}}$; otherwise, more weight is assigned to
$\widehat{\theta}_B^{\mathrm{MC}}$. \\

The same combined estimator can also be derived using classical control variate arguments \citep{ripley1987stochastic, bratley1987guide, hammersley1964monte}. Control variates reduce the variance of a MC estimator by leveraging known values of one or more related integrals \citep{Owen2000135}. Specifically, $h_1(\Y)$ can be treated as a control variate for $h_B(\Y)$, exploiting the fact that we assume $\mathbb{E}\!\left[h_1(\Y)\right] = \mathbb{E}[C \mid \aprior]$ to be known.
The resulting Monte Carlo estimator coincides with the convex combination estimator
$\widehat{\theta}_{\alpha}^{\mathrm{MC}}$, and the variance-minimizing coefficient
$\alpha^*$ is given by the classical control variate formula. 
As a consequence, VoI-formulation~(\ref{VoIa}) can then be seen as a control variate extension of formulation~(\ref{VoIb}), with a suboptimal coefficient of one. 
According to the control variate formulation \citep{asmussen2007stochastic}, the estimator $\widehat{\theta}_{\alpha}^{\mathrm{MC}}$ with optimized $\alpha^*$ reduces the variance of $\widehat{\theta}_B^{\mathrm{MC}}$
by the factor $\left(1 - \rho^2 \right)$ with 
$$\rho = \mathrm{Corr}(h_1,h_B) = \frac{\mathrm{Cov}(h_1(\Y), h_B(\Y))}{\sqrt{\mathrm{Var}(h_1(\Y)) \mathrm{Var}(h_B(\Y))}},$$
thus a $|\rho|$ closer to 1 leads to the best performance of $\widehat{\theta}_{\alpha}^{\mathrm{MC}}$. Furthermore, if the computational cost per replication is approximately unchanged by the use of the control variate, then $(1-\rho^2)$ quantifies the resulting computational speed-up \citep{glasserman2004monte}. \\

In practice, we have to rely on an estimator $\widehat{\alpha^*}$  based on the empirical variance and covariance employing samples $\{ \y_1, \dots \y_n \} \sim f_{\Y}(\y)$, denoted as $\widehat{\alpha_n}$. When the samples in use are the same as in the final estimator, the estimator is no longer unbiased \citep[][chapter~8]{owen2013}. Therefore, we also consider a bias-corrected variant based on the sample-splitting approach of \citet{AvramidisWilson1993}. The sample of size $n$ is partitioned into two equally sized subsets. A control variate coefficient, $\widehat{\alpha}^{*}_{n/2}$, is estimated from each subset and subsequently applied to the complementary subset when estimating the VoI. The resulting estimator given by the average is unbiased and is denoted by $\widehat{\theta}_{\alpha|2}^{\mathrm{MC}}(n)$.

\subsection{Stratified sampling}
\label{sec_lhs}
As a further variance–reduction technique, we consider stratified sampling. Stratified sampling partitions the sample space into mutually exclusive subsets, or strata, and allocates samples to each stratum in a controlled manner \citep{glasserman2004monte}. Stratified sampling is most effective when the variance of $\Y$ within each stratum is small \citep{glasserman2004monte}. Here, we focus on Latin hypercube sampling \citep[LHS][]{mckay1979comparison}.
LHS applies grid-based stratification to each one-dimensional marginal distribution and randomly combines the resulting marginal samples. This produces a more uniform coverage of each marginal distribution than standard MC sampling, while preserving unbiasedness of the resulting estimators \citep[see, e.g.,][]{asmussen2007stochastic}.
Because the stratification is applied independently to each marginal distribution, LHS is most effective when the integrand is dominated by first-order effects.
Using an LHS design with samples $\y_j \overset{\mathrm{LHS}}{\sim} f_{\Y}$, the corresponding estimators are the same as in Equation~\ref{eq:mc-estimators}, and we denote them by $\widehat{\theta}_A^{\mathrm{LHS}}(n)$ and $\widehat{\theta}_B^{\mathrm{LHS}}(n)$.

\subsection{Importance sampling}
\label{sec_is_gen}

For both expressions (\ref{VoIa}) and (\ref{VoIb}) of the VoI, we can estimate the integrals of interest using importance sampling (IS):
\begin{align}
\label{is_def}
        &\widehat{\theta_A^{\mathrm{IS}}} (n) = \frac{1}{n} \sum_{j=1}^{n} 
         \frac{f_{\Y}(\y_j) }  {g_{\Y}^A(\y_j)}
             h_A(\y_j) , \text{  with  } \y_j \sim g_{\Y}^A \\
        &\widehat{\theta_B^{\mathrm{IS}}} (n) = \mathbb{E}[C | \aprior] - \frac{1}{n} \sum_{j=1}^{n} 
         \frac{f_{\Y}(\y_j) }  {g_{\Y}^B(\y_j)}
             h_B(\y_j)  , \text{  with  } \y_j \sim g_{\Y}^B, \nonumber
        \end{align}   
with $g_{\Y}^A(\y) > 0$ whenever $f_{\Y}(\y) \; h_A(\y) \neq 0$ and $g_{\Y}^B(\y) > 0$ whenever $f_{\Y}(\y) \; h_B(\y) \neq 0$. The optimal IS densities $g_{\Y}^{A^*}$ and $g_{\Y}^{B^*}$ are proportional to \citep{kahn1953methods}:
    \begin{align}
        \label{is_perf}
        g_{\Y}^{A^*}: \quad \y \mapsto f_{\Y}(\y) \; | h_A(\y) | &= f_{\Y}(\y) \;\left[\mathbb{E}[C \mid \aprior, \y] - \min_{a} \; \mathbb{E}[C \mid a, \y]  \right] \\
        &= f_{\Y}(\y) \;\left[c_A(\aprior) + c_F \; p_F(\aprior | \y ) - \min_{a} \{ c_A(a) + c_F \; p_F(a | \y ) \}  \right], \nonumber \\
        g_{\Y}^{B^*}: \quad \y \mapsto f_{\Y}(\y) \; | h_B(\y) | &= f_{\Y}(\y) \; \min_{a} \; \mathbb{E}[C \mid a, \y] \nonumber \\
        &= f_{\Y}(\y) \; \min_{a} \{ c_A(a) + c_F \; p_F(a | \y ) \}. \nonumber
    \end{align} 

When the optimal densities are known, both estimators achieve zero variance. In practice, however, these ideal densities are not available and must be approximated, with the effectiveness of the method critically depending on the quality of the approximation. 

\subsubsection{Importance sampling density approximation}
We employ an approximation of the optimal IS densities (Eqs.~\ref{is_perf}), relying on a Gaussian process (GP) surrogate for the posterior failure
probabilities $\y \mapsto p_F(a | \y )$. GPs are a powerful and flexible tool for probabilistic modeling of unknown functions \citep[for details about GP modeling, we refer to ][]{roustant2012dicekriging}.  \\

We consider a finite action space
\(
\mathcal{A} = \{a^0, a^1, \dots, a^J\}.
\)
For each action \(a^j \in \mathcal{A}\), we model the mapping
\(
\y \mapsto \log p_F(a^j \mid \y)
\)
using a GP. Thereby, working on the logarithmic scale transforms the training probabilities to non-positive real‑valued training points. The resulting GP predictions are real-valued and have to be truncated after the back transform to ensure appropriate probabilities. We also considered a probit transform \citep{rasmussen2006gaussian}, but preliminary results suggest that the log transform performs better, even though the truncation is needed. 
To train the GP surrogates, we use an initial set of $n_0$ measurement outcomes
$\y_1,\dots,\y_{n_0}$ and corresponding log‑failure probabilities
$p_i^j = \log p_F(a^j \mid \y_i)$ for $j = 1,\dots, J$. We obtain $J$ GP surrogates for the mappings corresponding to each action.
Using the predictive distribution of the GP model in $\y$,
$$\log p_F(a^j \mid \y) \sim \mathcal{N}\!\left(\mu^j_n(\y), \sigma^j_n(\y)^2\right),$$
the failure probabilities $p_F(a^j \mid \y)$ follow a log‑normal distribution and we use the closed form of the expectation as approximation of the values:
$$ p_F(a^j \mid \y) \approx \min \left(1, \exp \left( \mu^j_n(\y) + \tfrac{1}{2}\sigma^j_n(\y)^2 \right) \right).$$ 
As noted earlier, we truncate the value from above at 1 to ensure appropriate probability values. 
This yields the following approximation of the expected cost for a given measurement $\y$ and action $a^j$:
\begin{align}
\label{cost_gp}
    \mathbb{E}[C \mid a^j, \y]
    &\approx
    c_A(a^j)
    + c_F \;
    \min \left(1, \exp \left( \mu^j_n(\y) + \tfrac{1}{2}\sigma^j_n(\y)^2 \right) \right).
\end{align}

Based on this approximation, we select the posterior‑optimal action given $\y$ and get
$\min_{a} \mathbb{E}[C \mid a, \y]$,
which yields an approximation of the optimal IS density $g_{\Y}^B$ (Eqs.~\ref{is_perf}).
For the IS density associated with estimator~(\ref{VoIa}), we additionally
require $\mathbb{E}[C \mid \aprior, \y]$, which can be approximated analogously using Equation~\ref{cost_gp} and the GP for $\aprior$. \\

This approach follows loosely the idea of \citet{dubourg2013metamodel} or \citet{dalbey2014gaussian}, who also perform an IS approach based on a GP surrogate in the context of rare event probability estimation.

\subsubsection{Importance sampling for control variate estimator}
\label{sec_is_cv}
The control variate estimator (Eq.~\ref{eq:mc-estimators-comb}) can be augmented by IS, as discussed by \citep{Owen2000135}. Therefore, we define the controlled importance sampling estimator as
\begin{align}
    \label{is_alpha}
    \widehat{\theta}_{\alpha}^{\mathrm{IS}}(n)
    &= (1-\alpha)\,\mathbb{E}[C \mid \aprior]
    + \frac{1}{n} \sum_{j=1}^n
    \frac{f_{\Y}(\y_j) }  {g_{\Y}^{\alpha}(\y_j)}
    \left[
        \alpha\, h_1(\y_j) - h_B(\y_j)
    \right], \text{ with } \y_j \sim g_{\Y}^{\alpha}, 
\end{align}
for which the best IS density $g_{\Y}^\alpha$ is proportional to $\y \mapsto f_{\Y}(\y) \; |\alpha\, h_1(\y) - h_B(\y)|$. However, this choice does not lead to a zero--variance estimator in general \citep[][chapter~9]{owen2013}.
The reason is that the integrand
$\alpha\, h_1(\y) - h_B(\y)$ may change sign over the support of~$\Y$. Nevertheless, the resulting density can still provide substantial variance reduction. \\

\subsubsection{Importance sampling mixtures}
Finally, we consider two more IS estimators mixing both estimators \ref{VoIa} and \ref{VoIb}, by splitting the number of samples into $n_A + n_B = n$, with allocating $n_A$ samples to the IS density $g_{\Y}^A$ and $n_B$ samples to the IS density $g_{\Y}^B$:
\begin{align}
\label{is_comb}
        \widehat{\theta_{AB}^{\mathrm{IS}}} (n) =
        \frac{n_A}{n} \widehat{\theta_A^{\mathrm{IS}}} (n_A)  + \frac{n_B}{n} \widehat{\theta_B^{\mathrm{IS}}} (n_B).
\end{align}  
As a heuristic to choose the numbers, we rely on the estimated $\widehat{\alpha^*}$ introduced for the control variate estimator (Eq. \ref{alpha}). We use both a gradual and a binary weighting strategy.
For the gradual estimator $\widehat{\theta_{A+B}^{\mathrm{IS}}}$, we choose
$n_A$ as the integer rounded value $\lfloor n \; \widehat{\alpha^*} \rceil $ and $n_B = n - n_A$.
For the binary strategy in~$\widehat{\theta_{A|B}^{\mathrm{IS}}}$, we set $n_A = n, n_B = 0$ if $\widehat{\alpha^*} \geq 0.5$, and vice versa.

\subsection{Sampling space}
In practice, $f_{\Y}$ is often not available in closed form. Instead of approximating $f_{\Y}$, as done in \citet{straub2014value}, we shift the analysis to the $\X$-space, where we assume that the density $f_{\X}$ is known or can be evaluated. Assuming the forward map $\mathcal{G}$ is deterministic and the measurement noise has PDF $f_{\Noise}$ and is independent of $\X$, 
we move all sampling approaches to the $\left(\X, \Noise \right)$-space. As an example, the MC estimates become: 
\begin{align*}
    \widehat{\theta}_A^{\mathrm{MC}}(n)
    = \frac{1}{n} \sum_{j=1}^n h_A(\mathcal{G} \left( \x_j \right) + \noise_j), \quad
    \widehat{\theta}_B^{\mathrm{MC}}(n)
    = \mathbb{E}[C \mid \aprior]
    - \frac{1}{n} \sum_{j=1}^n h_B(\mathcal{G} \left( \x_j \right) + \noise_j),
\end{align*}
with $\{ \x_1, \dots \x_n \} \sim f_{\X}(\x)$ and $\{ \noise_1, \dots \noise_n \} \sim f_{\Noise}(\noise)$. 

\section{Numerical illustrations}
\label{sec_test}

\subsection{Binary decision with perfect action}
\label{set_bin}

In our test cases, we consider a scenario in which the decision is between taking no action and taking a perfect action. This corresponds to a perfect maintenance/mitigation scenario, in which taking an action can completely eliminate the failure risk. Accordingly, the action space is
\(
\mathcal{A} = \{a^0, a^1\},
\)
where $c_A(a^0) = 0$ corresponds to zero cost for taking no action, and
$c_A(a^1) > 0$ corresponds to the cost of taking the action.
The action being perfect implies that both
\(
p_F(a^1) = 0
\)
and
\(
p_F(a^1 \mid \y) = 0
\)
for all $\y$ for which $f_{\Y}(\y) > 0$.
This leads to simplified expressions for the expected costs:
\[
\mathbb{E}[C \mid a^0] = c_F \, p_F(a^0), \qquad
\mathbb{E}[C \mid a^0, \y] = c_F \, p_F(a^0 \mid \y), \qquad
\mathbb{E}[C \mid a^1, \y] = \mathbb{E}[C \mid a^1] = c_A(a^1).
\]

\subsection{Implementation}

The R-code for the numerical illustrations of this study is available on GitHub: \newline (\href{https://github.com/LeaFrie/LastExpectation}{https://github.com/LeaFrie/LastExpectation}). Table~\ref{tab:methods} gives an overview of all the sampling methods. 
All methods are implemented in an underlying Gaussian space obtained by an iso-probabilistic transformation that preserves the probability content of the prior distribution.  
To approximate the IS density, we employ a GP model fitted with the \texttt{DiceKriging} package \citep{dicekriging_package}. For the approximation of the optimal IS density in the $\X$ space, we first analyze prior samples to identify the components of $\X$ that have the strongest influence on $\Y$. Importance sampling is then applied only to these dimensions, while the remaining dimensions are sampled using LHS.
The normalizing constant of the resulting IS density is computed numerically on a grid. 
To obtain stratified samples in the standard Gaussian space, the unit-interval samples generated within each stratum are transformed via the standard normal inverse CDF.
All LHS samples are generated using the \texttt{lhs} package \citep{lhs_package}. In general, we use the function \texttt{randomLHS}. For the dimensions selected for importance sampling, however, the initial design used to train the Gaussian process is generated with \texttt{maximinLHS} to achieve a more uniform spread of points across the input space. For $\widehat{\theta}_{\alpha}^{\mathrm{IS}}$, $\widehat{\alpha}_{n_0}$ is estimated from an initial LHS sample also used to construct the approximate optimal IS density. For $\widehat{\theta}_{A+B}^{\mathrm{IS}}$ and $\widehat{\theta}_{A|B}^{\mathrm{IS}}$, $\widehat{\alpha}_{n_0}$ is additionally used to determine the allocation between $n_A$ and $n_B$.  \\

For the estimation of the failure probabilities, we consider MC sampling:
\begin{equation}
    \widehat{p}_F = \frac{1}{n} \sum_{i=1}^n \mathbbm{I} \left\{ \mathcal{R}\left(\boldsymbol{x}_i, a^0 \right) \geq \thres \right\},
\end{equation}
with samples $\x_i \sim f_{\X}$ for $\widehat{p}_F(a^0)$ and $\x_i \sim f_{\X \mid \Y}$ for $\widehat{p}_F(a^0 \mid \y)$. Posterior samples are obtained from the post-burn-in states of a Markov chain Monte Carlo algorithm, with thinning applied to reduce autocorrelation and yield approximately independent samples. In general, we use $n = 10^6$ for $\widehat{p}_F(a^0)$ and $n = 8000$ for $\widehat{p}_F(a^0 \mid \y)$, the reduction being due to the increased computational effort of posterior sampling. Since the probabilities of interest may be very small, plain MC sampling with $n=8000$ can become unreliable: the coefficient of variation of the MC estimator already exceeds $5\%$ once $\widehat{p}_F \approx 0.05$ and reaches about $10\%$ once $\widehat{p}_F \approx 0.01$. We therefore adopt the following heuristic strategy: Whenever the plain MC estimate for $\widehat{p}_F(a^0 \mid \y)$ falls below $0.05$, we additionally evaluate it via posterior subset sampling \citep[as in][]{straub2016bayesian, friedli2024rare}. If the plain MC estimate lies between $0.01$ and $0.05$, we report the average of the two estimates; if it falls below $0.01$, where the plain MC estimate is too unreliable to be informative, we retain only the subset sampling estimate. We note that this switching rule is a pragmatic choice rather than a variance-optimal one; a more principled treatment could, e.g., combine both estimators through inverse-variance weighting. However, since the focus of this work lies elsewhere, we consider this heuristic sufficient to ensure reliable estimates of $\widehat{p}_F(a^0 \mid \y)$ across the full range of probabilities considered. \\

\begin{table}
    \centering
    \renewcommand{\arraystretch}{1.1}
    \begin{tabular}{llll}
        \textbf{Estimator} & \textbf{Notation} &  \textbf{Sample allocation} & \textbf{Estimated} \\
        && \textbf{($n$ in total)} & \textbf{parameters} \\
        \midrule
        MC (Eq. \ref{eq:mc-estimators}) & $\widehat{\theta_{A}^{\mathrm{MC}}}$, $\widehat{\theta_{B}^{\mathrm{MC}}}$   & $n$ from $f_{\Y}$ & - \\
        \midrule
        MC + control variate (Eq. \ref{eq:mc-estimators-comb})& $\widehat{\theta_{\alpha}^{\mathrm{MC}}}$, $\widehat{\theta_{\alpha|2}^{\mathrm{MC}}}$  & $n$ from $f_{\Y}$ & $\widehat{\alpha_n}$, $\widehat{\alpha_{n|2}}$ \\
        \midrule
        LHS (Eq. \ref{eq:mc-estimators})& $\widehat{\theta_{A}^{\mathrm{LHS}}}$, $\widehat{\theta_{B}^{\mathrm{LHS}}}$  & $n$ from $f_{\Y}$ with LHS & - \\
        \midrule
        IS (Eq. \ref{is_def} )& $\widehat{\theta_{A}^{\mathrm{IS}}}$, $\widehat{\theta_{B}^{\mathrm{IS}}}$ & $n_0$ from $f_{\Y}$,  & - \\ 
        && $(n- n_0)$ from $g_{\Y}^A$ / $g_{\Y}^B$ \\
        \midrule
        IS + control variate (Eq. \ref{is_alpha})& $\widehat{\theta_{\alpha}^{\mathrm{IS}}}$ &  $n_0$ from $f_{\Y}$,  & $\widehat{\alpha_{n_0}}$ \\
        && $(n- n_0)$ from $g_{\Y}^{\alpha}$ \\
        \midrule
        IS mixture (Eq. \ref{is_comb} ) & $\widehat{\theta_{A+B}^{\mathrm{IS}}}$ & $n_0$ from $f_{\Y}$, $n_A$ from $g_{\Y}^A$, & $\widehat{\alpha_{n_0}}$, \\
        && $(n-n_0-n_A)$ from $g_{\Y}^B$ & $n_A = \lfloor n \widehat{\alpha_{n_0}} \rceil $ \\
       \midrule
        IS mixture (Eq. \ref{is_comb})& $\widehat{\theta_{A|B}^{\mathrm{IS}}}$  & $n_0$ from $f_{\Y}$, $n_A$ from $g_{\Y}^A$, & $\widehat{\alpha_{n_0}}$,  \\
        && $(n-n_0-n_A)$ from $g_{\Y}^B$ & $n_A = n \lfloor  \widehat{\alpha_{n_0}} \rceil $
    \end{tabular}
    \caption{Overview of the sampling approaches used to estimate the VoI. Here, $\lfloor \cdot \rceil$ denotes rounding to the nearest integer.}
    \label{tab:methods}
\end{table}

\subsection{Test cases}

\subsubsection{Analytical example}
\label{sec_ana}

We present a proof of concept using a test case with an analytical solution. We consider a system with independent load $X_1$ and resistance $X_2$, with the vector $\mathbf{X} = (X_1, X_2)$ following a multivariate normal distribution with mean $(\mu_1, \mu_2)=(5.5,12)$ and variances $\sigma^2_1 = 4$ and $\sigma^2_2 = 3$. The quantity of interest is $\mathcal{R}(\mathbf{X}, a_0) = X_1 - X_2$, with failure defined as $\mathcal{R} > 0$. The measurement model is $\mathcal{G}(\mathbf{X}) = X_2$ with additive Gaussian noise of variance $\sigma_{\Noise}^2 = 0.01$. The binary action space assumes perfect maintenance with $p_F(a_1)=0$ and $p_F(a_1|\y)=0$, and the failure probabilities are given via the Gaussian CDFs:
\begin{equation*}
    p_F(a_0) = \Phi \left(\frac{\mu_1-\mu_2}{\sqrt{\sigma^2_1 + \sigma^2_2}} \right), \quad 
    p_F(a_0|\y) = \Phi \left( \frac{1}{\zeta} \left[ \mu_1 - 
    \frac{\sigma^2_2}{ \left(\sigma^2_2 + \sigma_{\Noise}^2 \right) }
    \y -  \frac{\sigma_{\Noise}^2}{\sigma_{\Noise}^2 + \sigma_2^2} \mu_2    \right] \right), 
\end{equation*}
where 
$\zeta = \sqrt{\frac{\sigma^2_2 \sigma_{\Noise}^2}{\sigma_2^2+\sigma_{\Noise}^2} + \sigma_1^2}$. We fix the cost of failure at $c_F = 10^6$ and test different maintenance costs $c_A(a_1)$. It holds that $p_F(a_0) = 0.007$, such that $c_A(a_1) \leq 7000$ results in $\aprior = a_1$, while $c_A(a_1) > 7000$ leads to $\aprior = a_0$. The resulting VoI increases sharply at low $c_A(a_1)$, reaches a maximum of about 4,900 at a level of $c_A(a_1) \approx 7000$, and then gradually decreases as the maintenance cost continues to rise. We consider different sampling spaces by also employing a version of the example with $\mathbf{X} = (X_1, X_2^1, \dots, X_2^{n_x})$, where $X_2^1, \dots, X_2^{n_x}$ are independent Gaussian random variables with variances $\frac{\sigma^2_2}{n_x}$. The total system resistance defined as $X_2 = X_2^1 + \dots + X_2^{n_x}$ is considered in $\mathcal{R}(\mathbf{X}, a_0)$ and $\mathcal{G}(\mathbf{X}) = X_2$, preserving the VoI values.  \\

\begin{figure}[p]
    \centering
    \includegraphics[width=.475\linewidth]{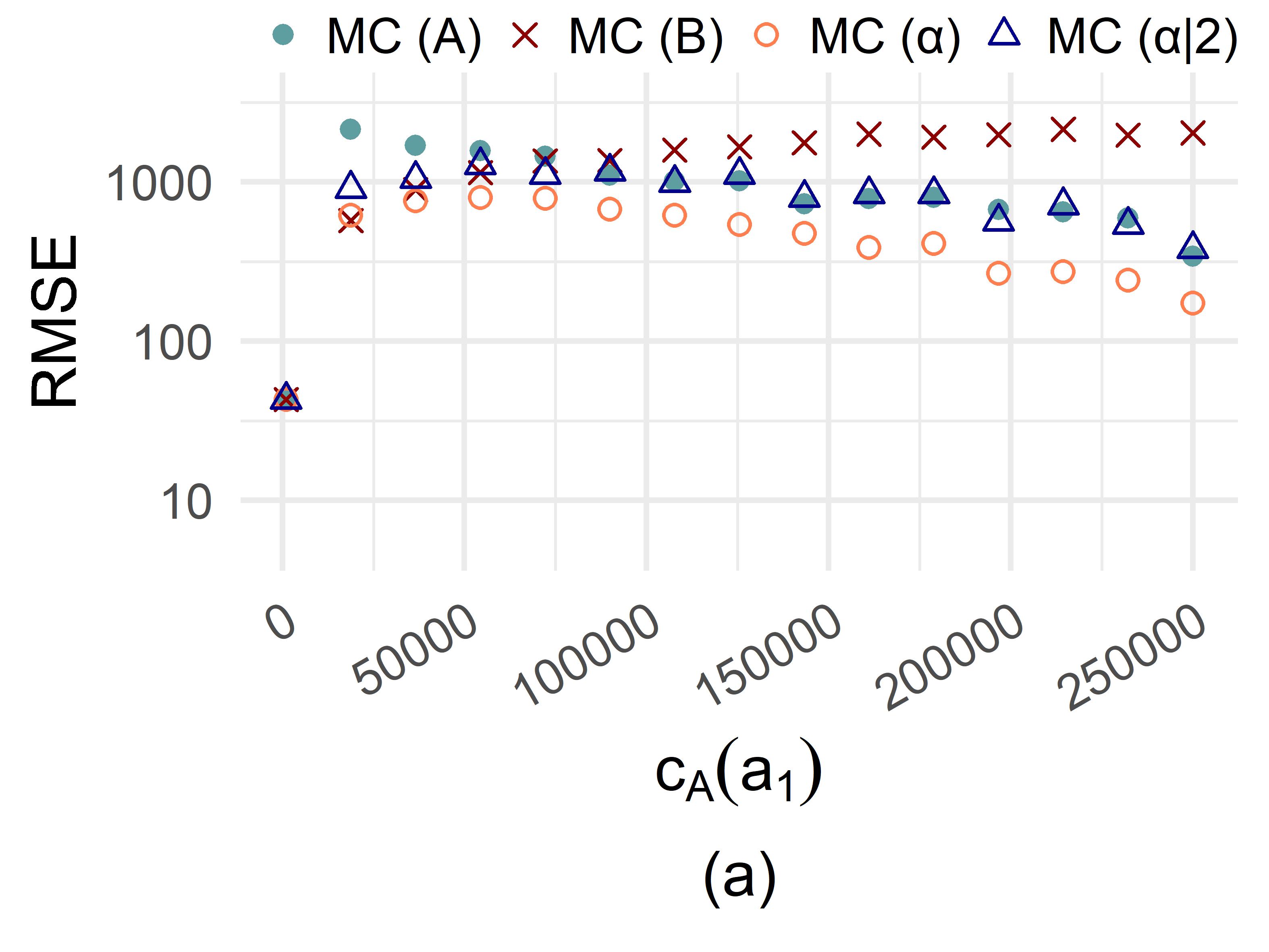}
    \includegraphics[width=.475\linewidth]{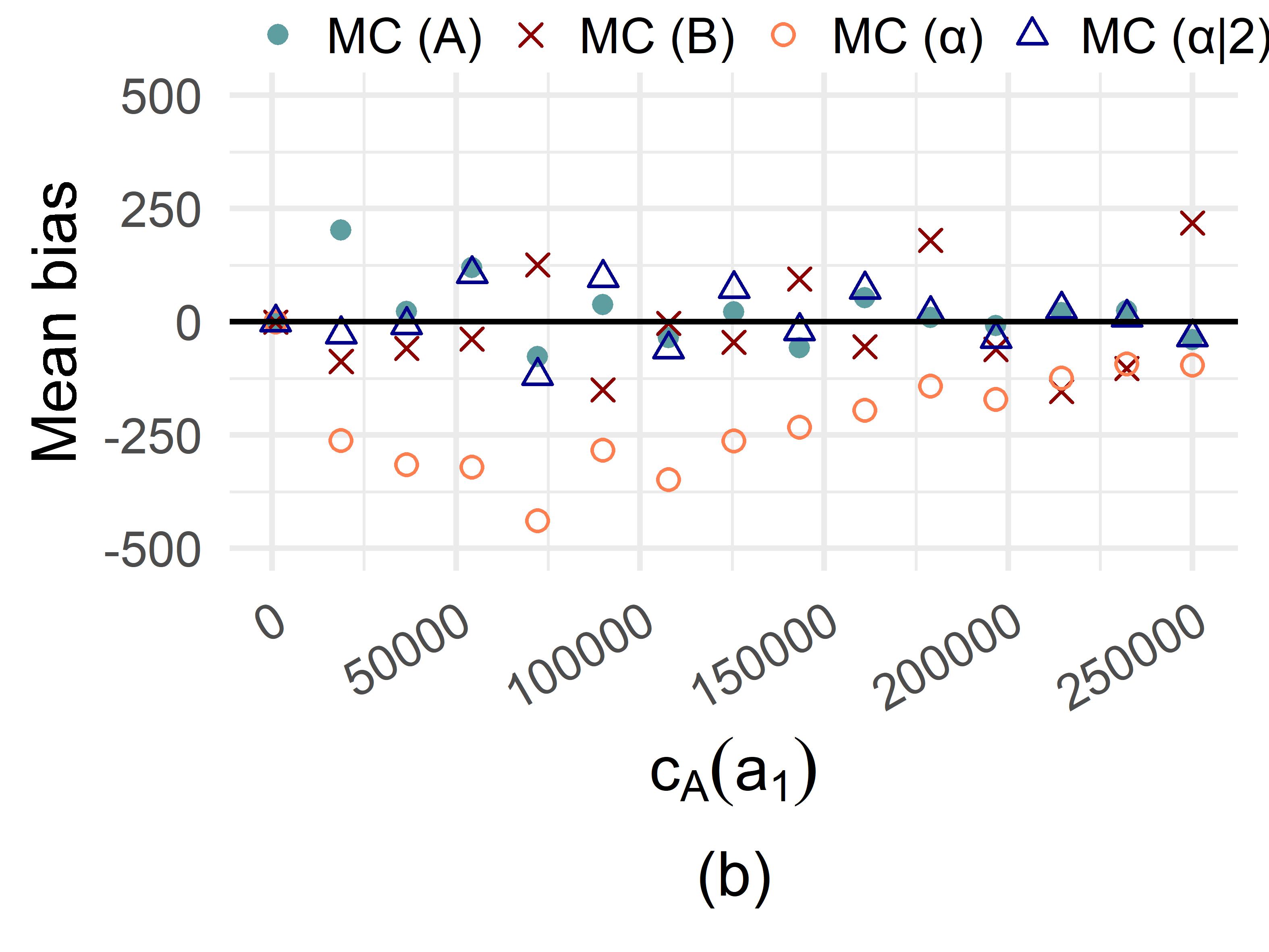} \\
    \includegraphics[width=.475\linewidth]{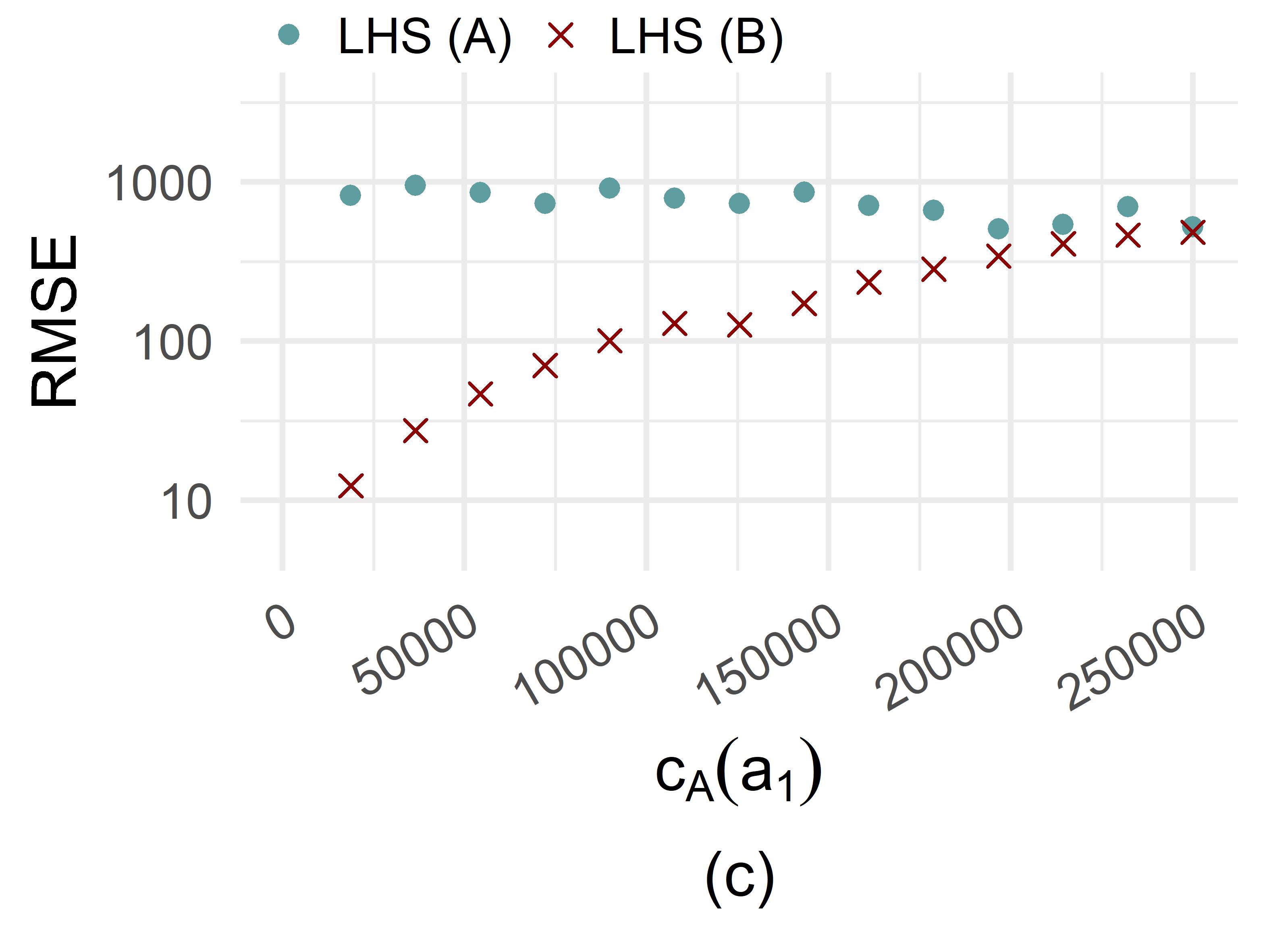}
    \includegraphics[width=.475\linewidth]{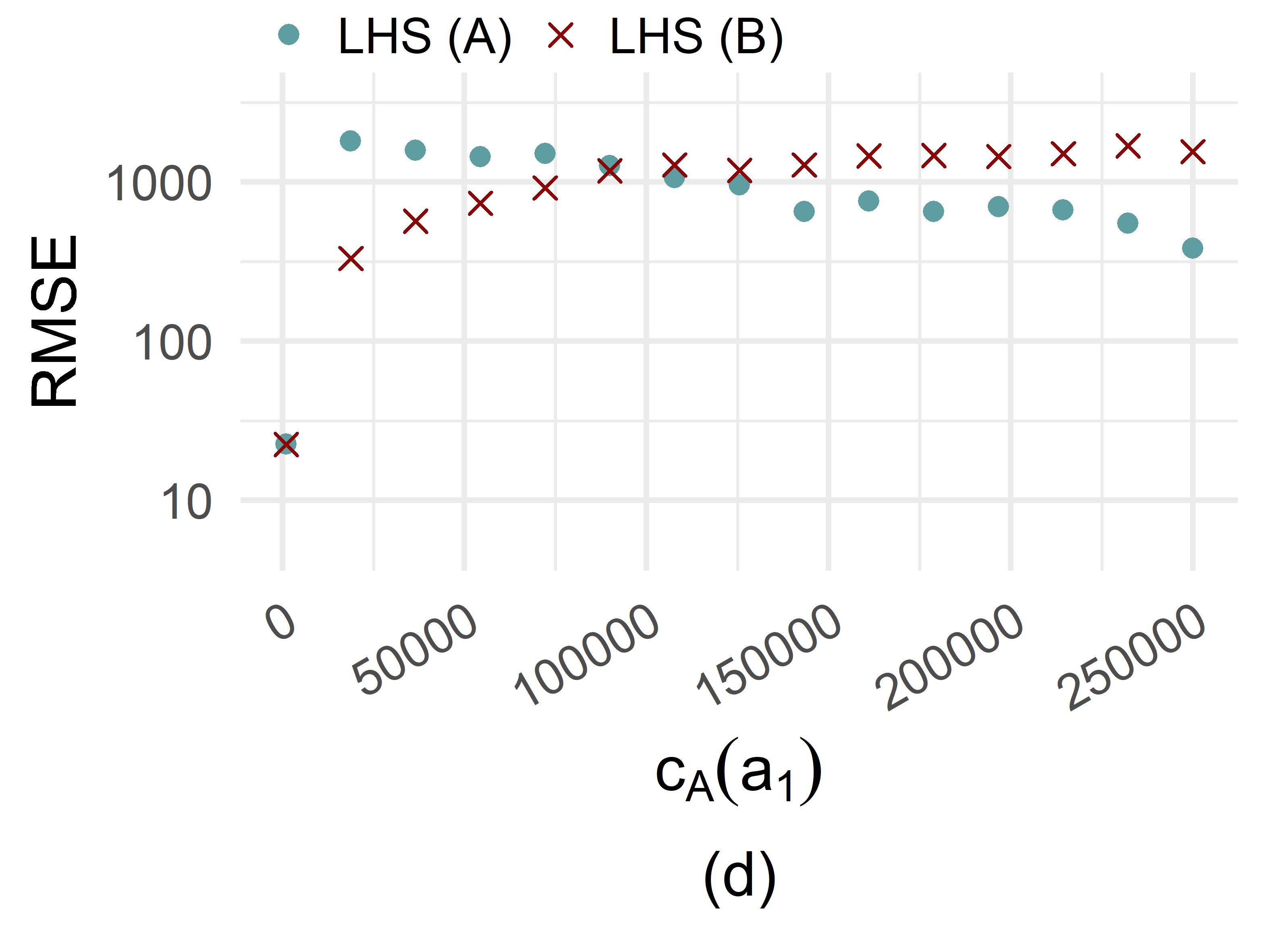} \\
    \includegraphics[width=.475\linewidth]{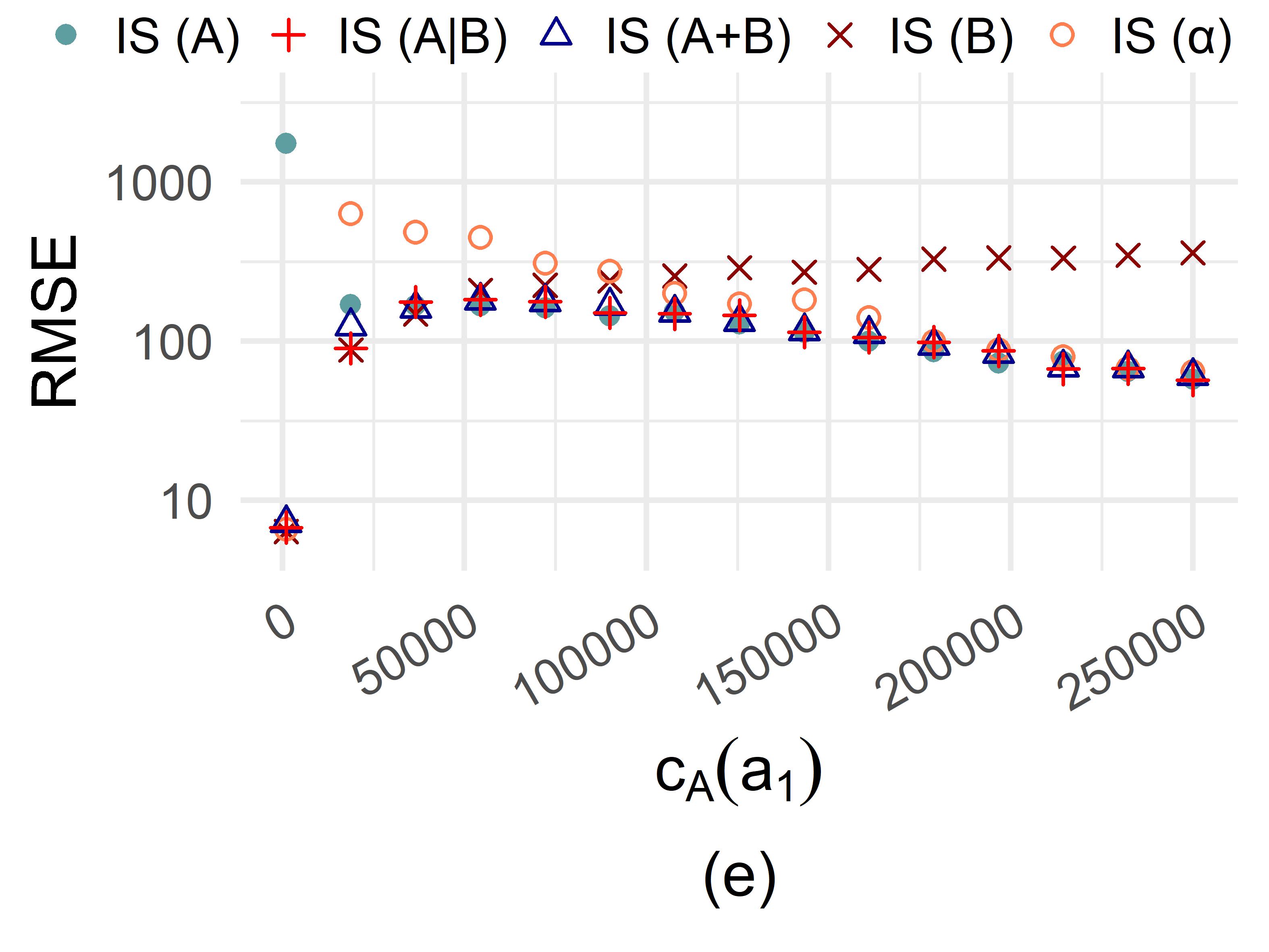}
    \includegraphics[width=.475\linewidth]{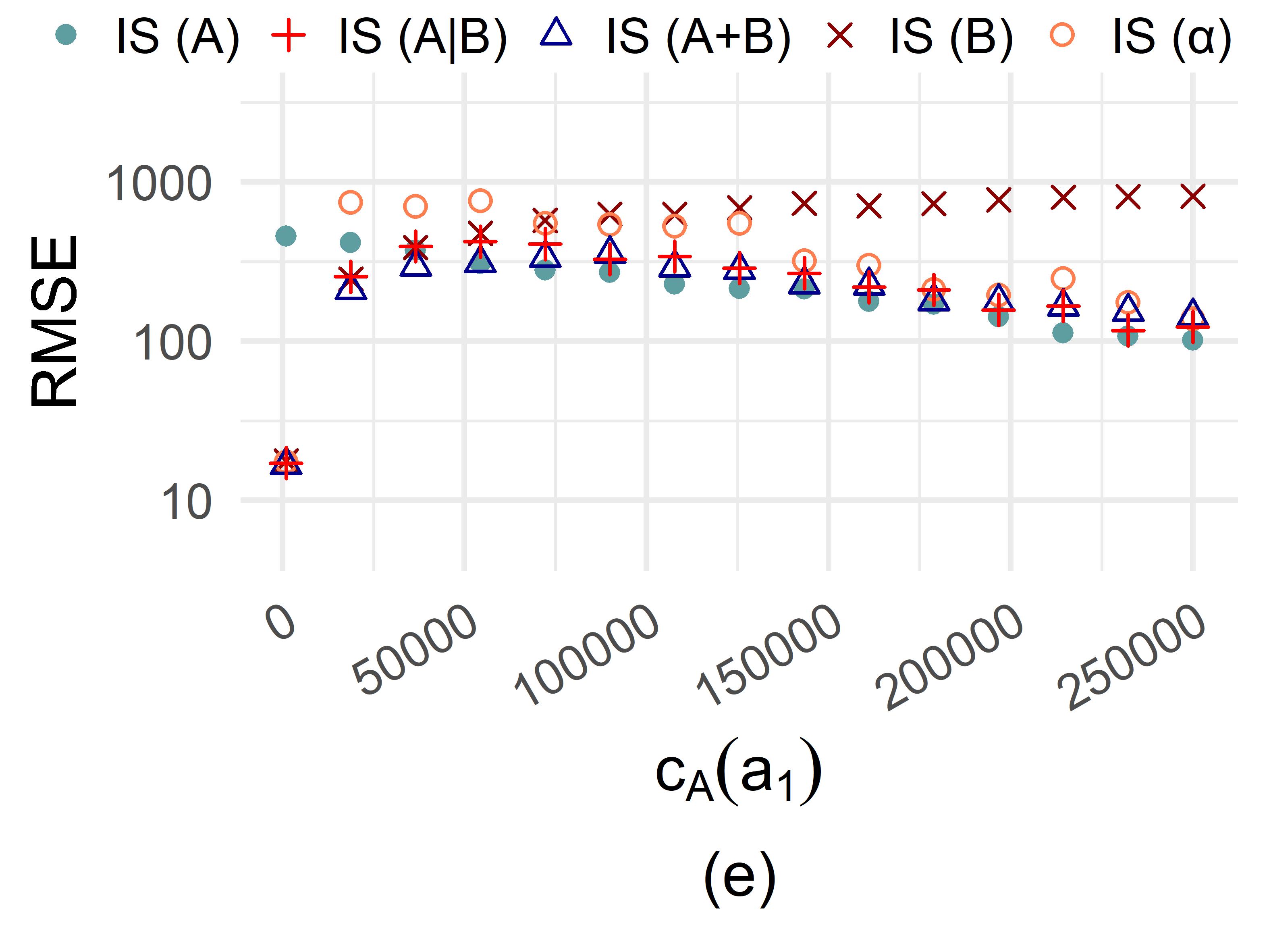}
    \caption{Analytical test example. Log-scale RMSE of the VoI estimators as a function of $c_A(a_1)$, over $250$ repetitions with $n = 100$ samples each. Sampling schemes: (a)~MC; (c)~LHS in $\mathbf{Y}$-space; (d)~LHS in $\mathbf{X}$-space with $n_x = 5$; (e)~IS in $\mathbf{Y}$-space with $n_0 = 30$; (f)~IS in $\mathbf{X}$-space with $n_x = 5$ and $n_0 = 50$. Panel (b) additionally shows the mean bias of the estimators under the different MC approaches.}
\label{fig:analytical_rmse} 
    \label{fig:ana_mc_lhs}
\end{figure}

\begin{figure}[h]
    \centering
    \includegraphics[width=.475\linewidth]{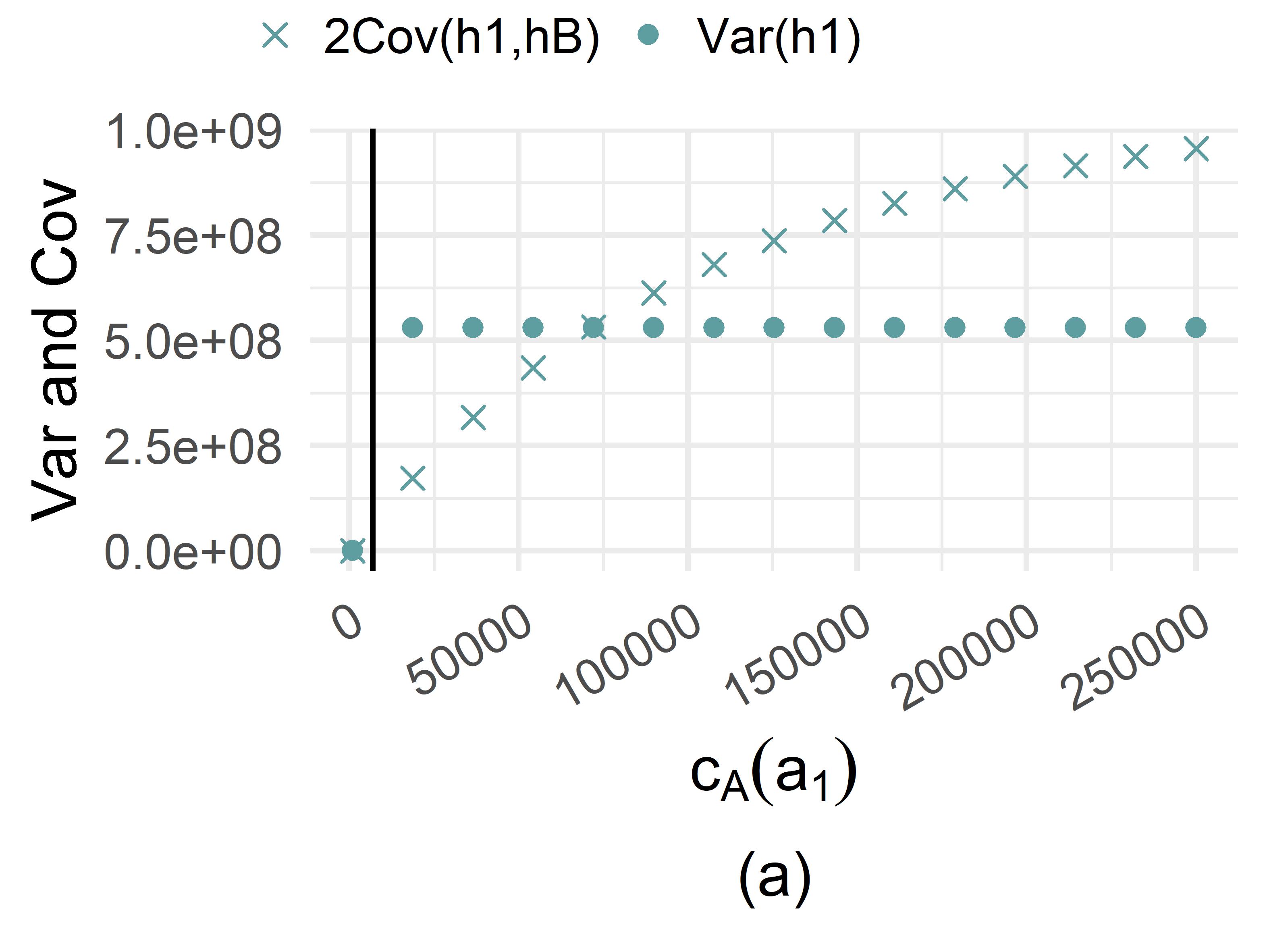}
    \includegraphics[width=.475\linewidth]{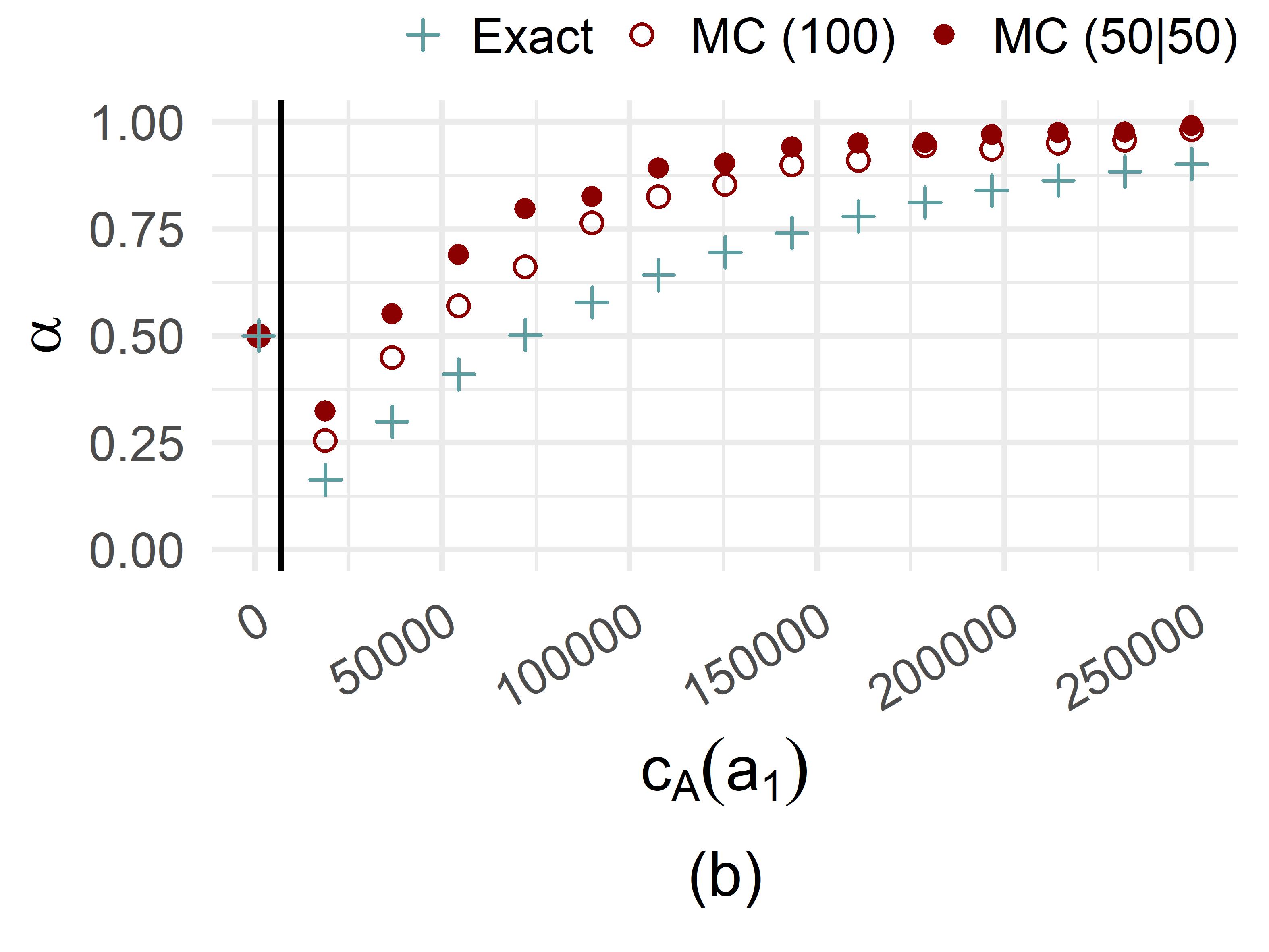}
    \caption{Analytical test example. (a) $\mathrm{Var}(h_1(\Y))$ and $2\,\mathrm{Cov}(h_1(\Y), h_B(\Y))$ for evolving $c_A(a_1)$ and (b) corresponding $\alpha^*$ values (in blue). The red dots in (b) depict the estimated $\alpha$ coefficients using $n=100$ MC samples as in $\widehat{\theta}_\alpha^{\mathrm{MC}}$ and the mean of the two coefficients as in $\widehat{\theta}_{\alpha|2}^{\mathrm{MC}}$ (mean over 250 repetitions). }
    \label{fig:ana_mc_lhs_cv}
\end{figure}

We compare the VoI estimates obtained with the different sampling methods against the reference values. Figure~\ref{fig:ana_mc_lhs} depicts the RMSE values for the different estimators, computed over 250 independent runs. 
In panel (a), we start by showing the results of the VoI estimation with $n=100$ MC samples. We observe that \(\widehat{\theta}_A^{\mathrm{MC}}\) performs better for larger \(c_A(a_1)\), while \(\widehat{\theta}_B^{\mathrm{MC}}\) performs better for smaller values. As expected, the control variate estimators \(\widehat{\theta}_\alpha^{\mathrm{MC}}\) and \(\widehat{\theta}_{\alpha|2}^{\mathrm{MC}}\) consistently match or outperform \(\widehat{\theta}_B^{\mathrm{MC}}\), with \(\widehat{\theta}_\alpha^{\mathrm{MC}}\) achieving lower RMSE values. However, panel (b), which shows the mean bias across the 250 repetitions, reveals a systematic negative bias in \(\widehat{\theta}_\alpha^{\mathrm{MC}}\), as discussed in Sect.~\ref{sec_cv}. The corrected estimator \(\widehat{\theta}_{\alpha|2}^{\mathrm{MC}}\) effectively eliminates this bias. Figure~\ref{fig:ana_mc_lhs_cv}(a) shows the relationship between $\mathrm{Var}(h_1(\Y))$ and $2\,\mathrm{Cov}(h_1(\Y),h_B(\Y))$ (see Sct.~\ref{sec_mc}). 
The point at which the MC estimators exchange their relative performance in Figure~\ref{fig:ana_mc_lhs}(a) is consistent with the intersection of $\mathrm{Var}(h_1(\Y))$ and $2\,\mathrm{Cov}(h_1(\Y),h_B(\Y))$. The estimated values of $\alpha^*$ for the control variate estimators can be seen in Figure~\ref{fig:ana_mc_lhs_cv}(b). We observe that MC estimation with $n = 50, 100$ systematically
overestimates the optimal coefficients $\alpha^*$, particularly for small and
moderate $c_A(a_1)$. This is not surprising, since the estimator of $\alpha^*$ is a
ratio of sample moments and therefore has an $\mathcal{O}(1/n)$ bias.
However, when the samples used to estimate the coefficients are independent of
those used in the final estimator, as in \(\widehat{\theta}_{\alpha|2}^{\mathrm{MC}}\),
the final estimator remains unbiased, although it is on average suboptimal. \\

Next, we are considering $n=100$ LHS samples in Figure~\ref{fig:ana_mc_lhs}(c) and (d). Focusing on the \(\Y\)-space (c, one-dimensional), the LHS estimator \(\widehat{\theta}_B^{\mathrm{LHS}}\) clearly improves the performance compared to MC. This is less pronounced for estimator \(\widehat{\theta}_A^{\mathrm{LHS}}\). For \(\X\)-space with \(n_x = 5\) depicted in panel (d), the improvement by LHS diminishes with the higher dimensional sampling space and the performance is similar to MC in panel (a). \\

\begin{figure}[h]
    \centering
    \includegraphics[width=.95\linewidth]{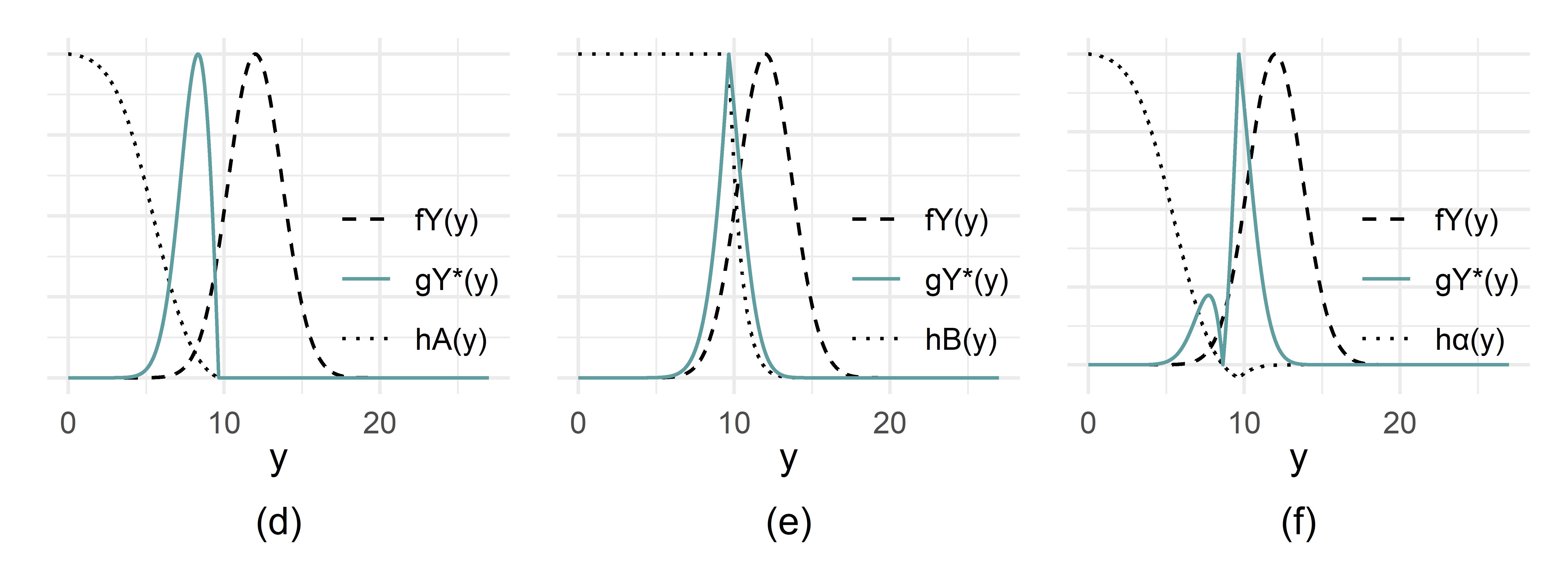}
    \includegraphics[width=.95\linewidth]{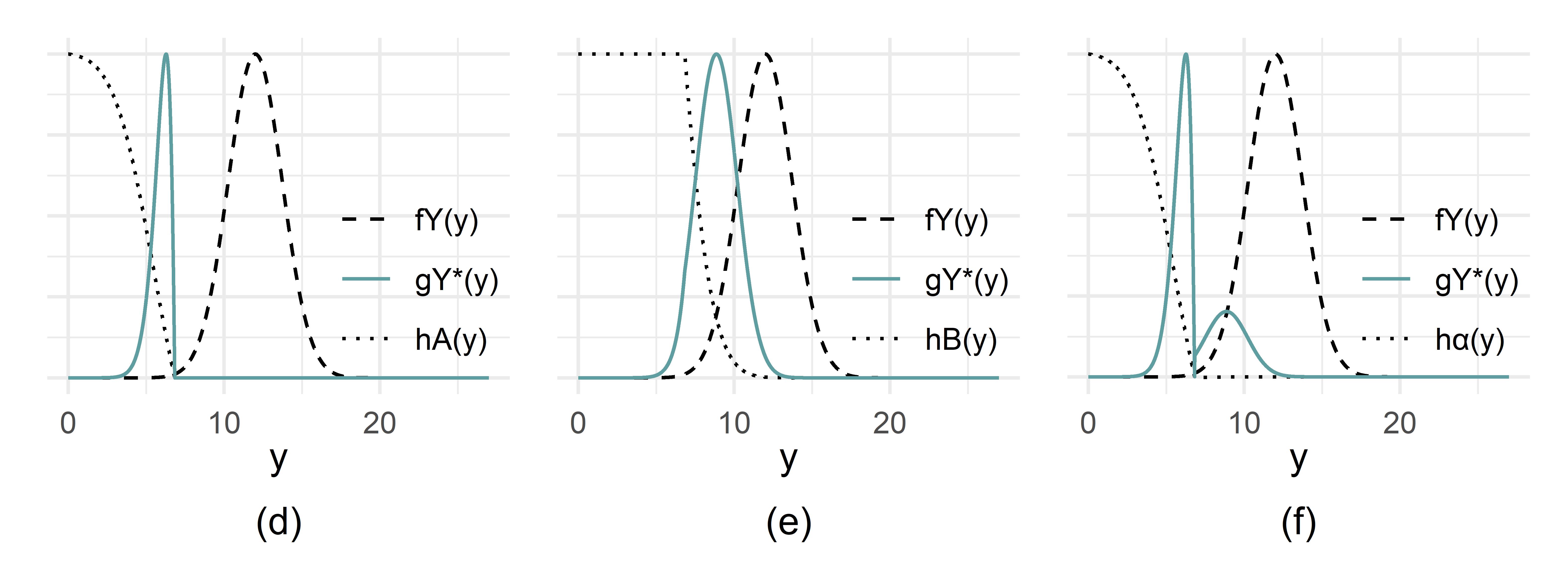}
    \caption{Analytical test example. 
    Optimal IS densities $g_{\Y}^*$ for (a,d) \(\widehat{\theta}_A^{\mathrm{IS}}\), (b,e) \(\widehat{\theta}_B^{\mathrm{IS}}\) and (c,f) \(\widehat{\theta}_\alpha^{\mathrm{IS}}\). The first row refers to \(c_A(a_1) = 11,600 \) and the second row to \(c_A(a_1) = 150,000 \). The dotted line $h_\alpha$ in the third column depicts $\alpha\, h_1(\y) - h_B(\y)$ (without taking the absolute value). }
    \label{fig:ana_optimalIS}
\end{figure}

Next, we are considering IS sampling. Figure~\ref{fig:ana_optimalIS} illustrates the optimal IS densities $g_{\Y}$ associated with the estimators $\widehat{\theta}_A^{\mathrm{IS}}$, $\widehat{\theta}_B^{\mathrm{IS}}$, and $\widehat{\theta}_\alpha^{\mathrm{IS}}$, shown in panels (a,d), (b,e), and (c,f), respectively. The top row corresponds to the case $c_A(a_1)=11{,}600$, whereas the bottom row corresponds to $c_A(a_1)=150{,}000$. In the third column, the dotted curve labeled $h_\alpha$ represents $\alpha h_1(\y)-h_B(\y)$ (without taking the absolute value). Using the absolute value in the optimal IS density leads to the bumps in $g_{\Y}$ observed in panels (c) and (f). In this analytical example, closed-form expressions for the optimal importance sampling densities can be derived, yielding IS estimators with zero variance. In practical applications, however, such exact solutions are generally unavailable. This motivates the use of a GP to approximate the optimal density. \\

Given the expected increase of the failure probability with decreasing resistance $X_2$, we employ a GP with linear trends in the input variables. 
The RMSEs of the GP-based IS approaches are shown in Fig.~\ref{fig:ana_mc_lhs}(e,f).
Given a total sampling budget of $n=100$, a trade-off must be made between the number of samples used for GP training and those reserved for the actual estimator. Good performance was obtained when approximately $30\%$ of the budget was allocated to GP training in the one-dimensional case, and $50\%$ in the five-dimensional case.
Panel~(e) shows the RMSE values for sampling in the $\Y$-space. In this one-dimensional setting, the LHS estimator $\widehat{\theta}_B^{\mathrm{LHS}}$ outperforms IS for smaller values of $c_A(a_1)$, whereas most IS approaches achieve lower RMSEs for larger values of $c_A(a_1)$. Panel~(f) presents the corresponding results in the $\X$-space for $n_x=5$. Here, IS clearly outperforms both LHS and MC. For example, at $c_A(a_1)=250{,}000$, the RMSE of LHS is reduced from 398 to 122 by using IS (with $\widehat{\theta}_{A|B}^{\mathrm{IS}}$). 
Regarding the basic IS estimators, the results mirror those observed for the MC estimators: 
$\widehat{\theta}_B^{\mathrm{IS}}$ performs better for smaller values of $c_A(a_1)$, while 
$\widehat{\theta}_A^{\mathrm{IS}}$ is preferable for larger values. Overall, the sampling-allocation estimators $\widehat{\theta}_{A+B}^{\mathrm{IS}}$ and 
$\widehat{\theta}_{A|B}^{\mathrm{IS}}$ show consistently good performance. In contrast, $\widehat{\theta}_\alpha^{\mathrm{IS}}$ performs poorly for smaller values of $c_A(a_1)$ due to the absolute-value issue discussed in Section~\ref{sec_is_cv}. This is also evident 
from Fig.~\ref{fig:ana_optimalIS}(c,f). For smaller $c_A(a_1)$ values, the untransformed quantity $h_\alpha(\y)=\alpha h_1(\y)-h_B(\y)$ is negative over much of the region where $f_{\Y}$ has substantial mass, resulting in more frequent sign changes and increased variance, even under the optimal IS density.

\subsubsection{Fatigue reliability}
\label{sec_crack}
We investigate the fatigue reliability example of \citet{kanjilal2023bayesian}, which was adapted from \citet{ditlevsen1996structural}, A fatigue crack is modeled starting from an initial crack size $q_0$. Crack growth follows Paris' law,
\[
q(d) = \left[\left(1-\frac{b}{2}\right)C\Delta S^b\pi^{b/2}d + q_0^{1-b/2}\right]^{\frac{1}{1-b/2}},
\]
where $d$ denotes the number of load cycles, $\Delta S$ the constant stress range, and $C$ and $b$ are empirical parameters. 
We define failure by $q(8\times10^5) > q_c$, with $q_c=50$ mm. The prior distributions are given by $q_0\sim\mathrm{Exp}(1)$, $\Delta S\sim\mathcal{N}(60,10^2)$, and $(\ln C,b)$ following a bivariate normal distribution with mean $(-33,3.5)$, standard deviations $(0.47,0.3)$, and correlation coefficient $\rho_{\ln(C),b}=-0.9$. 
We consider a single crack-size measurement $y$ at $d=10^5$, with additive Gaussian measurement noise $\sigma_{\Noise}=0.2$ and likelihood
\[
L(\x)=\exp\!\left[-\frac{1}{2}\left(\frac{\mathcal{G}(\x,d)-y}{\sigma_{\Noise}}\right)^2\right].
\]
The failure cost is set to $c_F=10^6$, while for the maintenance cost we consider two values, $c_A(a_1)=10^4$ and $c_A(a_1)=10^5$. \\

\begin{figure}
    \centering
    \includegraphics[width=\linewidth]{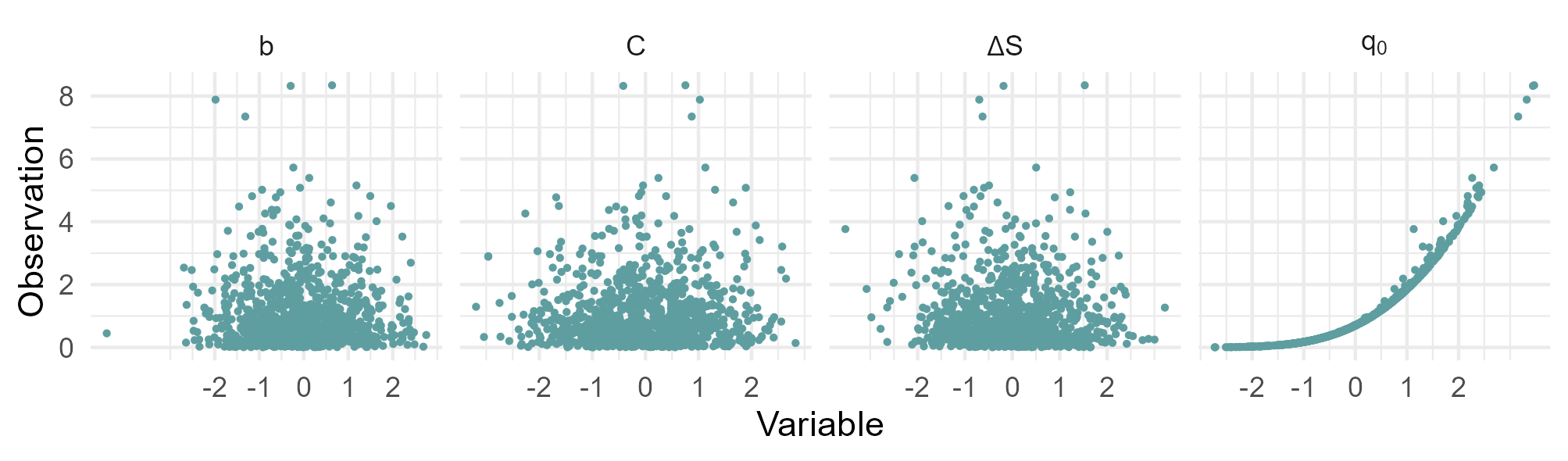}
    \caption{Fatigue crack growth. Scatter plot of the measurement $\y$ versus the the different input parameters (1000 MC samples). }
    \label{fig:crack1}
\end{figure}

We start by determining the prior decision $\aprior$ and the corresponding prior cost $\mathbb{E}[C | \aprior]$. To this end, we rely on $10^6$ MC samples, obtaining a (prior) probability of failure of $9.217 \times 10^{-3}$ and $\mathbb{E}[C | \aprior] = 9,217$. For both considered values of \( c_A(a_1) \), this leads to $\aprior = a_0$.
In this example we do not have the option of working in the $\Y$-space as $f_{\Y}$ is not available. Before fitting an IS density for all four $\X$-dimensions, we use 1000 of the MC samples drawn before and generate the corresponding measurement outcomes, which are shown in Figure~\ref{fig:crack1}. The corresponding first-order Sobol indices \citep[also estimated based on 1000 MC samples using the method of ][]{monod2006uncertainty} are 0.98 for $q_0$ and 0.06-0.07 for $C$, $\Delta S$ and $b$. Apparently, the measurement $\y$ (and thus the change in failure probability from  prior to posterior) is primarily sensitive to $q_0$. Therefore, the IS density is constructed only for this variable, while the remaining dimensions are sampled using LHS. Given the expected monotonic increase of the crack-size measurement (and thus failure probability) with increasing initial crack size $q_0$, we employ a GP with a linear trend. \\

Table~\ref{tab:crack} summarizes the mean values and standard deviations (SDs) of the VoI estimates obtained from 100 repetitions with $n=100$. Again, we use 30 \% of the sampling budget to train the GP-based IS density.
While for $c_A(a_1)=10^4$ all methods yield comparable mean estimates, for $c_A(a_1)=10^5$ the mean of the estimator $\widehat{\theta_{\alpha}^{\mathrm{MC}}}$ is clearly lower than that of the other estimators, suggesting a bias introduced by the reuse of the same samples (Sect.~\ref{sec_cv}); this bias disappears with $\widehat{\theta_{\alpha|2}^{\mathrm{MC}}}$.
For $c_A(a_1)=10^4$, the lowest SDs are obtained with $\widehat{\theta}_B^{\mathrm{LHS}}$, $\widehat{\theta}_B^{\mathrm{IS}}$, and $\widehat{\theta}_{A|B}^{\mathrm{IS}}$. 
By contrast, $\widehat{\theta}_A^{\mathrm{MC}}$ and $\widehat{\theta}_A^{\mathrm{LHS}}$ perform poorly.
On the other hand, for $c_A(a_1)=10^5$, the smallest SDs are achieved by $\widehat{\theta}_A^{\mathrm{IS}}$, $\widehat{\theta}_{A+B}^{\mathrm{IS}}$ and $\widehat{\theta}_{A|B}^{\mathrm{IS}}$, with values in the same range as for $c_A(a_1)=10^4$.
Here, $\widehat{\theta}_A^{\mathrm{MC}}$, $\widehat{\theta}_B^{\mathrm{MC}}$ and $\widehat{\theta}_A^{\mathrm{LHS}}$ perform the worst. 
Considering both scenarios, the estimator $\widehat{\theta}_{A|B}^{\mathrm{IS}}$ exhibits comparably stable performance across both cases.

\begin{table}[h!]
\centering
\renewcommand{\arraystretch}{1.5}

\begin{tabular}{
    l c|
    c c|
    c c
}
 & & \multicolumn{2}{c}{$c_A(a_1)=10^4$}
   & \multicolumn{2}{c}{$c_A(a_1)=10^5$} \\

Method 
 & Samples $n$
 & Mean & SD & Mean & SD \\
 \hline
 $\widehat{\theta_{A}^{\mathrm{MC}}}$ & 100 & 5523 & 2722  & 952 & 1821 \\
 $\widehat{\theta_{B}^{\mathrm{MC}}}$ & 100 &  5913 & 377  &  1309 & 1726 \\
 $\widehat{\theta_{\alpha}^{\mathrm{MC}}}$ & 100 &  5700 &  438 & 541  & 684 \\

 $\widehat{\theta_{\alpha|2}^{\mathrm{MC}}}$ & 100  & 5793  & 551  &  939 & 1427 \\

 \hline
 $\widehat{\theta_{A}^{\mathrm{LHS}}}$ & 100 & 5656 & 1643 & 1136 & 1702 \\
 $\widehat{\theta_{B}^{\mathrm{LHS}}}$ & 100 &  5776 &  126 & 1098 & 526 \\

\hline
 $\widehat{\theta_{A}^{\mathrm{IS}}}$ & 100 ($n_0 = 30$) &   5643 & 670  &   974 & 188  \\
 $\widehat{\theta_{B}^{\mathrm{IS}}}$ & 100 ($n_0 = 30$)  &  5822 & 217  &  1159 & 522 \\
 $\widehat{\theta_{\alpha}^{\mathrm{IS}}}$ & 100 ($n_0 = 30$)  &  5818 & 280 &   986 & 384 \\
 $\widehat{\theta_{A+B}^{\mathrm{IS}}}$ & 100 ($n_0 = 30$)  &  5807 & 357  & 1013 & 301  \\
 $\widehat{\theta_{A|B}^{\mathrm{IS}}}$ & 100 ($n_0 = 30$)  & 5822 & 217 & 1008 & 275 
\end{tabular}
\caption{Fatigue reliability: Mean and standard deviation (SD) of the VoI estimates obtained with the different sampling approaches over 100 repetitions. }
\label{tab:crack}
\end{table}

\subsubsection{Slope stability}
\label{sec_slope}

As a third test case, we consider the reliability of a saturated infinite clay slope under undrained conditions \citep{kanjilal2023bayesian}. The slope has height $H=5$~m, inclination $\beta=15^\circ$, and unit weight $\gamma=20$~kN/m$^3$. The factor of safety at depth $d$ is defined as
\[
FS(d) = \frac{s_u(d)}{\gamma d \sin \beta \cos \beta},
\]
and failure occurs if $\min_d FS(d) < 1$. The undrained shear strength is modeled as
\[
s_u(d) = s_{u0} + b \gamma d \exp[w(d)],
\]
where $w(d)$ is a zero-mean Gaussian random field with $\sigma_w=0.24$ and exponential auto-correlation function with 
correlation length $1.9$~m, represented via a truncated Karhunen–Loève expansion with six Gaussian variables $x_1^{KL}, \dots, x_6^{KL}$. The parameters $s_{u0}$ and $b$ follow lognormal priors with means $14.67$~kPa and $0.272$, and standard deviations $4.04$~kPa and $0.189$, respectively. The soil column is discretized into 100 layers. A single measurement of $s_u$ at 2~m depth is considered, assuming multiplicative lognormal measurement noise \citep{kanjilal2023bayesian}. The costs are $c_F=10^6$ and $c_A(a_1)=2\times 10^5$ or $c_A(a_1)=9\times 10^5$. \\

\begin{figure}
    \centering
    \includegraphics[width=\linewidth]{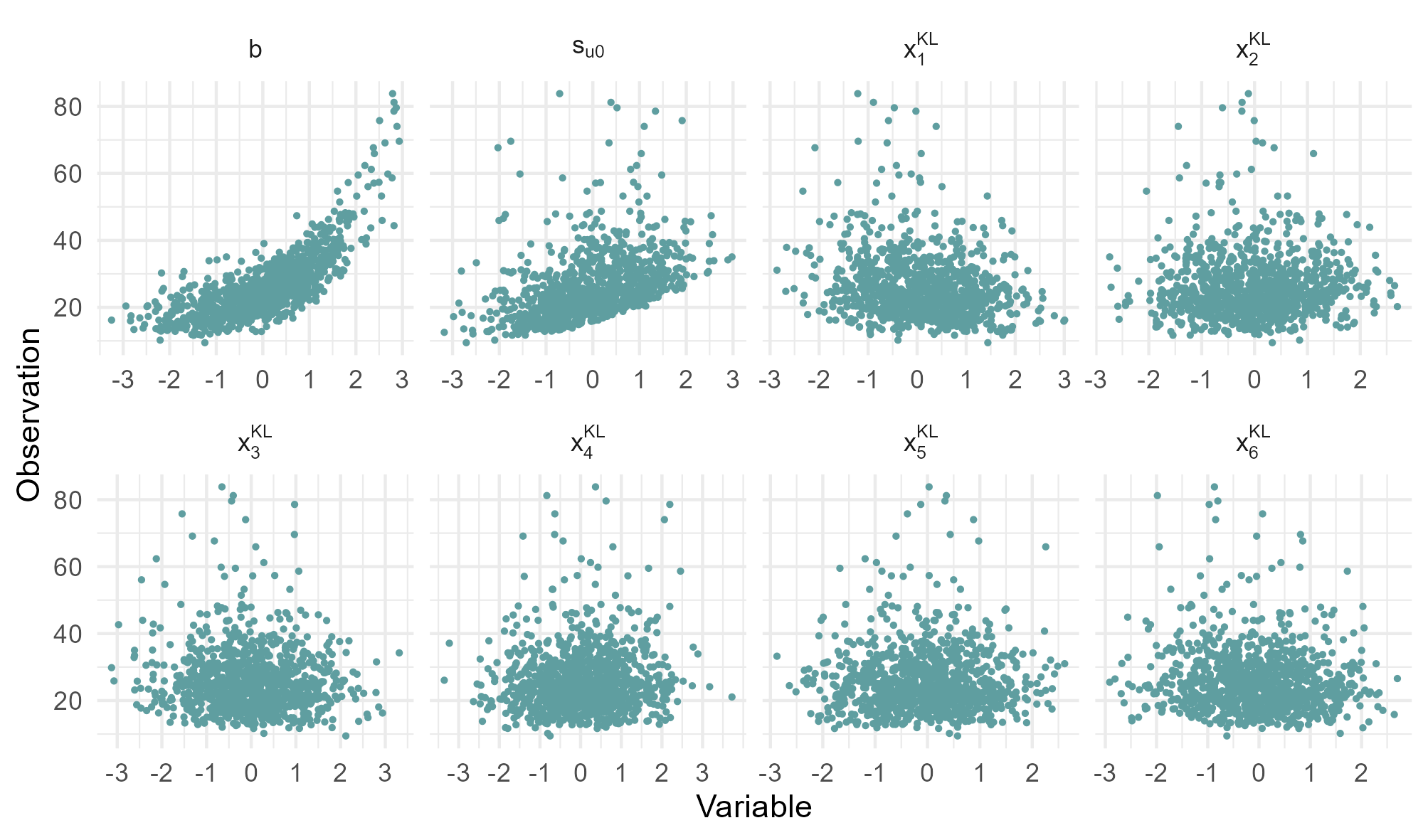}
    \caption{Slope stability. Scatter plot of $\y$-values versus the the different input parameters (1000 MC samples). }
    \label{fig:slope1}
\end{figure}

Again, we begin with a prior analysis, yielding a prior failure probability of 0.146 and an expected prior cost of $\mathbb{E}[C\mid\aprior]=146{,}000$, estimated using $10^6$ MC samples. Consequently, the prior-optimal decision is $\aprior=a_0$ for both mitigation costs.
Figure~\ref{fig:slope1} illustrates the sensitivity of the measurement to the input parameters based on 1000 MC samples. The corresponding Sobol indices, also estimated from 1000 MC samples, are 0.18 for $s_{u0}$, 0.71 for $b$, 0.04 for $x_1^{\mathrm{KL}}$, and below 0.01 for all remaining variables. Since the measurement is predominantly influenced by $b$ and $s_{u0}$, we test importance sampling with those variables, while the remaining dimensions are sampled using LHS. 
Given the expected monotonic behavior of shear strength (and thus the failure probability) as a function of $b$ and $s_{u0}$, we again employ a GP model with linear trends in both variables.\\

As for the previous test case, Table~\ref{tab:slope} shows the mean values and SDs of the VoI estimates obtained from 100 repetitions with $n=100$.
All methods produce means within a similar range for both scenarios; the deviating means can be explained by the high standard deviation of the corresponding estimates.
Considering the `pure' estimators with MC, LHS and IS, we observe very large SD when using $\widehat{\theta}_{A}^{\mathrm{MC}}$ and $\widehat{\theta}_{A}^{\mathrm{LHS}}$ for $c_A(a_1)=2 \times 10^5$, and when using $\widehat{\theta}_{B}^{\mathrm{MC}}$ and $\widehat{\theta}_{B}^{\mathrm{LHS}}$ for $c_A(a_1)=9 \times 10^5$. For MC sampling, the control variate estimators $\widehat{\theta}_{\alpha}^{\mathrm{MC}}$ and $\widehat{\theta}_{\alpha \mid 2}^{\mathrm{MC}}$ perform well across both scenarios, with hardly any difference when using the bias correction.
Using IS, the best results across both $c_A(a_1)$ are obtained with $\widehat{\theta}_{\alpha}^{\mathrm{IS}}$ and $\widehat{\theta}_{A|B}^{\mathrm{IS}}$. 
However, an improvement compared to $\widehat{\theta}_{\alpha|2}^{\mathrm{MC}}$ was only achieved for $c_A(a_1)=9 \times 10^5$.

\begin{table}[h!]
\centering
\renewcommand{\arraystretch}{1.5}

\begin{tabular}{
    l c|
    c c|
    c c
}
 & & \multicolumn{2}{c}{$c_a(a_1) = 2 \times 10^5$}
   & \multicolumn{2}{c}{$c_a(a_1) = 9 \times 10^5$ }\\

Method 
 & Samples $n$ 
 & Mean & SD & Mean & SD  \\

 \hline
 $\widehat{\theta_{A}^{\mathrm{MC}}}$ & 100 & 78791 & 19634 & 1174 & 1028 \\
 
 $\widehat{\theta_{B}^{\mathrm{MC}}}$ & 100 & 81598 & 9118 & 3981 & 26392 \\
 
 $\widehat{\theta_{\alpha}^{\mathrm{MC}}}$ & 100 & 80171 &  4760  & 1105 & 899 \\

 $\widehat{\theta_{\alpha|2}^{\mathrm{MC}}}$ & 100 &  80456 &  4764   & 1142 & 919 \\

 \hline
 $\widehat{\theta_{A}^{\mathrm{LHS}}}$ & 100 &  81498 & 11362  & 1353 &  864 \\
 
 $\widehat{\theta_{B}^{\mathrm{LHS}}}$ & 100 & 80684 & 4368  & 539 & 12966 \\


\hline
 $\widehat{\theta_{A}^{\mathrm{IS}}}$ & 100 ($n_0 = 30)$  & 81403 & 14294 &  1275 & 632   \\
 
 $\widehat{\theta_{B}^{\mathrm{IS}}}$ & 100 ($n_0 = 30)$ & 80726 & 6820  &  -1489 & 25652  \\
 
 $\widehat{\theta_{\alpha}^{\mathrm{IS}}}$ & 100 ($n_0 = 30)$  &   79122 & 5002 &  1285 &  710    \\
 
 $\widehat{\theta_{A+B}^{\mathrm{IS}}}$ & 100 ($n_0 = 30)$  & 79048 &  7503   &  1596 & 2032  \\
 
 $\widehat{\theta_{A|B}^{\mathrm{IS}}}$ & 100 ($n_0 = 30)$  & 80266 & 6820 &  1275 &  632 
 
\end{tabular}
\caption{Slope stability. Mean and SD of VoI estimates for each sampling approach (100 repetitions). }
\label{tab:slope}
\end{table}

\section{Discussion and outlook}

In all three test cases, we found that relying exclusively on VoI formulation (B), as is common in much of the existing VoI literature \citep[e.g.,][]{pozzi2011,zonta2014,straub2014value,dutta2019value,kamariotis2022value,LANGTRY2025126549}, is generally suboptimal. Likewise, relying solely on formulation (A) \citep[as in, e.g.,][]{jiang2020optimization} does not provide a robust alternative. The performance of both formulations varies considerably across scenarios, leading to inconsistent estimation accuracy.
This manuscript showed that formulation (A) can be seen as a control variate extension of formulation (B), with a (suboptimal) coefficient of one. When relying on the optimal coefficient, the performance of the resulting estimator becomes more stable across scenarios. \\ 

For MC-based estimation, the bias-corrected control-variate estimator $\widehat{\theta}_{\alpha|2}^{\mathrm{MC}}$ is recommended as it provides consistently robust performance across the considered test cases. LHS is particularly effective for low-dimensional sampling problems (Fig~\ref{fig:analytical_rmse}). However, for higher-dimensional spaces, such as the $\X$-space example in Sect.~\ref{sec_ana}, it provided only limited additional variance reduction compared with standard MC sampling. Additionally, the choice of estimator has a strong influence on performance, as demonstrated in Sect.~\ref{sec_slope}, where large differences were observed across scenarios and between VoI formulations (A) and (B).  \\

When effective importance sampling densities can be constructed, the estimator $\widehat{\theta}_{A|B}^{\mathrm{IS}}$ is generally a reliable option. However, the benefits of IS were problem-dependent: little improvement was observed in the slope stability example (Tab.~\ref{tab:slope}), whereas in the fatigue reliability example the standard deviation was more than halved relative to $\widehat{\theta}_{\alpha|2}^{\mathrm{MC}}$ in both scenarios (Tab.~\ref{tab:crack}). 
Considering Figures~\ref{fig:crack1} and \ref{fig:slope1}, together with the corresponding first-order sensitivity indices, the strategy of selecting variables for IS based on first-order effects appears to perform well in the fatigue reliability example, but is less effective in the slope stability example. This observation is also consistent with the strong performance of LHS in the former case, particularly for $c_A(a_1)=10^4$. In the slope stability example, a larger number of variables and higher-order interactions may contribute substantially to the quantity of interest, making it more challenging to construct an approximation of the optimal importance sampling density.
Furthermore, the benefits of importance sampling are most pronounced when changes in the optimal decision are driven by rarely observed values of $\Y$. This is illustrated by the crack growth example with $c_A(a_1)=10^5$, where the VoI is only about 1000, and IS achieves substantially greater variance reductions than either MC or LHS. \\

The proposed methods could be improved in several directions. 
In particular, the importance sampling approach could be refined by adaptively updating the surrogate model as new samples become available and by reusing the samples employed to construct the IS density in the final estimator. Furthermore, the current grid-based approximation of the normalizing constant could be replaced by more scalable approaches, such as sequential Monte Carlo methods. 
Furthermore, while we focused on Latin hypercube sampling, other stratified sampling techniques, such as adaptive stratification, optimal-allocation stratification, or orthogonal-array-based designs, may offer additional variance reductions and warrant further investigation. On another note, one could also consider a combined estimator relying on control variates and stratified sampling.
However, the estimation of the optimal coefficient has to be adapted with respect to the changed sampling and can be very hard to identify \citep[][chapter~9]{owen2013}. As an alternative to such a global coefficient, one could consider stratum-specific control variates or coefficients, as first discussed by \citet{grant1983hybrid}. While potentially beneficial, such approaches are often difficult to apply in practice. \\

Beyond these methodological refinements, the framework could also be extended to more general decision problems involving more than two actions or even continuous action spaces. While the IS approach, which relies on the decision framework of Section~\ref{sec:simple}, would require some adaptation, the control variate and LHS approaches are applicable to more general settings without modification. Furthermore, it would be interesting to consider problems with higher-dimensional or non-continuous observations. We have not focused on the computation of the failure probabilities. We assume that all failure probabilities can be evaluated with negligible uncertainty. However, if probability of failure evaluations are subject to centered noise, unbiasedness is preserved but the variance of the resulting estimates is increased. Finally, we have treated the two expectations appearing in the VoI separately, effectively assuming a two-step approach; these computations could also be addressed jointly \citep{straub2014value}.

\section{Conclusion}
    We investigated three classical variance-reduction techniques, namely control variates, stratified sampling, and importance sampling, as well as combinations thereof, for sampling-based estimation of the VoI, focusing on the expectation over possible measurement outcomes.
    Across all three considered test cases, relying exclusively on either of the two common VoI formulations proved suboptimal, as their performance varied considerably between scenarios and could lead to highly inconsistent results. By establishing a control-variate relationship between the two formulations, we derived more robust and efficient VoI estimators.
    Overall, the control-variate estimator based on MC sampling showed a consistently solid performance and represents a suitable default choice. Additional gains were obtained through stratified sampling in low-dimensional sampling spaces and through importance sampling when efficient proposal densities could be constructed, particularly in scenarios where decision changes are driven by rare measurement outcomes. While the proposed methods provide general guidance for efficient VoI estimation, considerable space remains for methodological improvements and extensions.

\paragraph{Acknowledgments} Lea Friedli acknowledges financial support from the Swiss National Science Foundation under grant number 225353.
We also acknowledge Dr. Habib Najm for pointing us in the direction of control variates.

\paragraph{Data availability statement}
No real-world data were used in this study and all numerical illustrations are based exclusively on synthetic data. The R code required to reproduce the numerical examples and results presented in this paper is publicly available on GitHub:  (\href{https://github.com/LeaFrie/LastExpectation}{https://github.com/LeaFrie/LastExpectation}). 

\paragraph{Declaration of generative AI and AI-assisted technologies in the manuscript preparation process}
During the preparation of this work the author(s) used the M365 Copilot based on the GPT-5 chat model in order to improve the clarity and quality of the writing, debug and optimize code and support brainstorming. After using this tool/service, the author(s) reviewed and edited the content as needed and take(s) full responsibility for the content of the published article.

\bibliographystyle{plainnat}  
\bibliography{ref}      

\end{document}